\documentclass{aa}

\usepackage{natbib}
\usepackage{hyperref}
\hypersetup{colorlinks=true,
            linkcolor=blue,
            citecolor=blue,
            filecolor=blue,
            urlcolor=blue,
            }

\usepackage{graphicx}	
\usepackage{amsmath}	
\usepackage{amssymb}	
\usepackage{multicol}        
\usepackage{bm}		
\usepackage{xcolor}
\usepackage{subcaption}
\usepackage[labelfont=bf]{caption}
\usepackage{stfloats}

\newcommand{\hGpc}{{\ifmmode{h^{-1}{\rm Gpc}}\else{$h^{-1}$Gpc}\fi}}
\newcommand{\hMpc}{{\ifmmode{h^{-1}{\rm Mpc}}\else{$h^{-1}$Mpc}\fi}}
\newcommand{\hkpc}{{\ifmmode{h^{-1}{\rm kpc}}\else{$h^{-1}$kpc}\fi}}
\newcommand{\hMsun}{{\ifmmode{h^{-1}{\rm {M_{\odot}}}}\else{$h^{-1}{\rm{M_{\odot}}}$}\fi}}
\newcommand{\Mstar}{{\ifmmode{M_{*}}\else{$M_{*}$}\fi}}
\newcommand{\Mhalo}{{\ifmmode{M_{\rm Halo}}\else{$M_{\rm Halo}$}\fi}}
\newcommand{\Ngal}{{\ifmmode{N_{\rm gal}}\else{$N_{\rm gal}$}\fi}}
\newcommand{\Msun}{{\ifmmode{{\rm {M_{\odot}}}}\else{${\rm{M_{\odot}}}$}\fi}}
\newcommand{\ltsima}{$\; \buildrel < \over \sim \;$}
\newcommand{\gtsima}{$\; \buildrel > \over \sim \;$}
\newcommand{\lsim}{\lower.5ex\hbox{\ltsima}}
\newcommand{\gsim}{\lower.5ex\hbox{\gtsima}}

\newcommand{\Tab}[1]{Table~\ref{#1}}
\newcommand{\Sec}[1]{Section~\ref{#1}}

\newcommand{\Fig}[1]{Fig.~\ref{#1}}

\usepackage[T1]{fontenc}
\usepackage{ae,aecompl}

\usepackage{newtxtext,newtxmath}
\usepackage{natbib}
\bibpunct{(}{)}{;}{a}{}{,} 

\begin{document} 

   \title{The NIVARIA-Local Group simulations:\\ Unveiling the origin of rotation curve diversity in dwarf galaxies and its connection to galaxy morphology}
   
\subtitle{}

\titlerunning{NIVARIA-LGs}

   \author{A. Contreras-Santos
          \inst{1,2}\fnmsep\thanks{E-mail: ana.contreras@iac.es}
          \and
          A. Di Cintio \inst{1,2}\fnmsep\thanks{E-mail: adicintio@iac.es}
          \and
          A. Negri \inst{3,1}
          \and
          J. Sarrato-Al\'os \inst{2,1}
          \and
          S. Cardona-Barrero \inst{1,2}
          \and \\
          G. García-Bethencourt \inst{1}
          \and
          C. Brook \inst{2,1}
          \and
          N. I. Libeskind \inst{4}
          \and
          A. Knebe \inst{5,6,7}
          \and
          A. V. Macci\`o \inst{8}
          }

   \institute{Departamento de Astrof\'{i}sica, Universidad de La Laguna, Av. Astrof\'{i}sico Francisco Sanchez s/n, 38206 La Laguna, Tenerife, Spain
         \and
             Instituto de Astrof\'{i}sica de Canarias, C/V\'{i}a L\'{a}ctea s/n, 38205 La Laguna, Tenerife, Spain
        \and
             Facultad de Física, Universidad de Sevilla, Avda. Reina Mercedes s/n, Campus de Reina Mercedes, E-41012 Sevilla, Spain
        \and 
            Leibniz Institut für Astrophysik Potsdam (AIP), An der Sternwarte 16, D-14482 Potsdam, Germany
        \and
            Departamento de F\'isica Te\'{o}rica, M\'{o}dulo 15, Facultad de Ciencias, Universidad Aut\'{o}noma de Madrid, 28049 Madrid, Spain
        \and
            Centro de Investigaci\'{o}n Avanzada en F\'isica Fundamental (CIAFF), Facultad de Ciencias, Universidad Aut\'{o}noma de Madrid, 28049 Madrid, Spain
        \and
            International Centre for Radio Astronomy Research, University of Western Australia, 35 Stirling Highway, Crawley, Western Australia 6009, Australia
        \and
            Center for Astrophysics and Space Science (CASS), New York University Abu Dhabi, PO Box 129188, Abu Dhabi, UAE
                     }

\date{Received September 30, 2026; accepted xxxxxxxxxxxxx}

 
  \abstract{
  Dwarf galaxies exhibit a large diversity in the shapes of their rotation curves, posing a long-standing challenge for models of galaxy formation within the $\Lambda$CDM paradigm. 
  We present here \texttt{NIVARIA-LG}, a new suite of  cosmological simulations for studying dwarf galaxy evolution in realistic Local Group environments. This intermediate-resolution pilot work lays the groundwork for the future, full-scale NIVARIA project. The simulations combine constrained Local Group initial conditions from \texttt{HESTIA} with the \texttt{NIHAO} galaxy formation model.
  The simulated galaxies reproduce observed stellar-to-halo mass relations, match the Baryonic Tully-Fisher relation, and capture the distribution of observed stellar effective radii, including compact and diffuse galaxies. 
  We analyse the structural and kinematic properties of isolated dwarf galaxies with $10^7 \lesssim M_\star/M_\odot \lesssim 10^{10}$, focusing on the origin of rotation curve diversity. \texttt{NIVARIA-LG}  naturally produces a broad range of rotation curve shapes, including slowly rising systems underrepresented in previous simulation suites, as well as steep rotation curves.
  We find that rotation curve diversity is primarily associated with gas extent and inner dark matter structure, with more extended gas and shallower inner density profiles corresponding to more slowly rising rotation curves, while stellar effective radius shows little independent association. 
  The evolution of individual galaxies further shows that rotation curve shapes  change rapidly as gas is redistributed during galaxy assembly. We also find that recent accretion events are associated with more slowly rising rotation curves in the most massive dwarf galaxies. Together, these results suggest that rotation curve diversity is connected to the evolution of the baryonic cycle, which in turn shapes the inner dark matter distribution of galaxies. The rich environment of the Local Group and its filaments influences this evolution by regulating gas accretion and redistribution, producing the observed diversity in rotation curve shapes.

  }


   \maketitle
%

\section{Introduction}

The $\Lambda$ cold dark matter ($\Lambda$CDM) paradigm has been remarkably successful in describing the large-scale structure of the Universe and the growth of cosmic structure across a wide range of scales. In this framework, galaxies form within dark matter halos through the interplay between gravitational collapse and baryonic processes. Dwarf galaxies occupy the low-mass end of the galaxy population and are among the most dark matter-dominated systems known. Their shallow gravitational potentials make them particularly sensitive to feedback, environmental effects, and other baryonic processes that regulate galaxy formation and evolution. As a result, dwarf galaxies provide valuable constraints on the interplay between dark matter and baryonic physics.

Despite its success on cosmological scales, the $\Lambda$CDM model has faced a number of challenges on smaller scales (less than $\sim1$ Mpc in length and $\sim10^{11}$ $M_\odot$ in mass, \citealp{Bullock-BK2017}). Discrepancies between theoretical predictions and observations of dwarf galaxies have led to a number of well-known challenges such as the missing satellites problem \citep{Klypin1999,Moore1999}, the too-big-to-fail problem \citep{Boylan-Kolchin2011,Boylan-Kolchin2012,Garrison-Kimmel2014}, or the cusp--core problem \citep{Flores1994,Moore1994,Burkert1995}. Although significant progress has been made in understanding the role of baryonic processes in alleviating many of these tensions, dwarf galaxies remain a critical testing ground for models of galaxy formation and the nature of dark matter. 

Among these challenges, one of the most puzzling ones is the large diversity observed in the rotation curves of dwarf galaxies \citep{Oman2015}. This is closely related to the cusp--core problem, which highlights the discrepancy between the cuspy Navarro-Frenk-White \citep[NFW;][]{Navarro1996,Navarro1997} density profiles predicted by dark matter-only simulations and the cored inner mass distributions inferred in many observed dwarf galaxies. However, while modern hydrodynamical simulations have demonstrated that baryonic feedback can significantly modify the inner structure of dark matter halos and alleviate this tension \citep[e.g.][]{Read2005,Pontzen2012}, reproducing the full range of observed rotation curve shapes remains a challenge for current models. 

Several simulation suites have explored this issue with differing outcomes. \citet{Oman2015} reported little diversity in the \texttt{EAGLE} simulations, which do not form prominent dark matter cores in dwarf-scale galaxies. In contrast, the \texttt{NIHAO} simulations, which do produce core formation over a range of masses \citep{DiCintio2014b}, are able to reproduce slowly rising rotation curves, although they still struggle to match the most steeply rising systems \citep{Santos-Santos2018}. Similarly, modified implementations of \texttt{EAGLE} that include more efficient core formation improve agreement for cored systems but continue to miss the most rapidly rising rotation curves \citep{Santos-Santos2020}. 

Modern cosmological simulations have made significant progress in addressing the observed diversity by exploring a wider range of baryonic and feedback processes. For instance, \citet{Koudmani2025} show that additional feedback channels such as AGN feedback can increase the spread of inner density profiles, while variations in star formation and feedback efficiency similarly broaden the predicted range of rotation curve shapes \citep{Cruz2025}. Despite these advancements, reproducing the systems with the most slowly rising rotation curves remains challenging \citep[see also][]{Jackson2025}. At the same time, part of the observed diversity may arise from observational and modelling systematics. Non-circular gas motions and deviations from idealised disc rotation can bias the inferred kinematics and artificially increase the scatter in rotation curve shapes \citep{Roper2023,Dado2026}. Overall, while progress is being made on both modelling and observational fronts, the origin of the full observed diversity remains uncertain, and disentangling the relative contribution of different effects is non-trivial. 

Addressing this problem requires simulations that simultaneously capture the internal structure of dwarf galaxies, the baryonic processes capable of modifying their dark matter distributions, and the environmental effects that may contribute to the observed diversity. To this end, we introduce the \texttt{NIVARIA-LG} suite (Numerical InVestigation of dwARf galaxies Including AGN feedback in the Local Group), a new set of cosmological zoom-in simulations designed to study the origin of dwarf galaxy diversity within a realistic Local Group environment.

\texttt{NIVARIA-LG} builds upon the constrained initial conditions of the \texttt{HESTIA} project \citep{Libeskind2020}, which successfully reproduces the large-scale structure and dynamics of the Local Group and its surroundings. By construction, these simulations provide a realistic environmental context for studying dwarf galaxy evolution, allowing environmental effects to be investigated alongside internal galaxy processes \citep[e.g.][]{Dupuy2022,Newton2023,Garcia-Bethencourt2026}. However, the galaxy formation model adopted in \texttt{HESTIA}, based on the Auriga simulations \citep{Grand2017}, produces predominantly cuspy dark matter density profiles in dwarf galaxies \citep[e.g.][]{Gannon2025}, as well as an excess of stars in the dwarf galaxy regime compared to observed stellar-to-halo mass relations. These limitations prevent \texttt{HESTIA} from fully exploring the role of baryon-driven core formation in shaping dwarf galaxy rotation curves.

To investigate this phenomenon, we resimulate the \texttt{HESTIA} initial conditions using the \texttt{Gasoline2} hydrodynamics code and galaxy formation model \citep{Wadsley2017}. This implementation has been shown to produce baryon-driven modifications of the inner dark matter density profiles of dwarf galaxies over a broad range of stellar masses, approximately $10^6 \lesssim M_\star/M_\odot \lesssim 10^{9.5}$ \citep{Governato2012,DiCintio2014a,Tollet2016}. By combining a realistic Local Group environment with a baryonic model capable of producing a broad range of inner density profiles, \texttt{NIVARIA-LG} provides a unique framework for studying the physical origin of dwarf galaxy diversity.

In this first paper of the series, we present the first, initial simulation of the \texttt{NIVARIA} suite and use it to investigate the structural and kinematic diversity of dwarf galaxies. Our primary goal is to identify the physical drivers of rotation curve diversity and to assess the relative importance of dark matter density structure, galaxy size, gas content, and environment in shaping the observed kinematics of low-mass galaxies. 
We describe the simulations setup and properties in \Sec{sec:simulations}. In \Sec{sec:galaxies} we present the galaxy population in the \texttt{NIVARIA-LG} simulation and how we compute different properties for these galaxies. Our results on the diversity of rotation curves and physical interpretation is shown in \Sec{sec:results}. Finally, in \Sec{sec:conclusions} we summarise our conclusions and discuss future prospects.

\section{The NIVARIA Simulations} \label{sec:simulations}

The NIVARIA project, acronym for `Numerical InVestigation of dwARf galaxies Including AGN feedback', is a large upcoming suite of galaxy simulations designed to investigate the impact of different feedback mechanisms on dwarf galaxies in a cosmological environment (PI A. Di Cintio). The project will include both cosmological Local Group simulations and controlled simulations of isolated dwarf galaxies, allowing the effects of feedback to be explored under different environmental conditions. 

In this paper, we present the first, pilot simulation of the Local Group suite, hereafter referred to as \texttt{NIVARIA-LG}. This initial run provides the baseline simulation for subsequent comparisons and does not include the black hole (BH) module or AGN feedback. Future work will use the same initial conditions and numerical framework to introduce the BH module and investigate the effects of AGN feedback, as well as to explore higher-resolution realisations.


The \texttt{NIVARIA-LG} simulation is a constrained cosmological simulation designed to reproduce the Local Group and its surrounding large-scale environment. It is a periodic box of side length 147.5 Mpc (100 $h^{-1}$Mpc), with a central zoomed-in sphere of radius 5 $h^{-1}$Mpc. The full box is initially run with dark matter only, while the zoomed region is run with full hydrodynamics with the SPH code \texttt{Gasoline2}. The mass resolution achieved in this region is $m_\mathrm{DM} = 1.6 \times 10^6\,M_\odot$, $m_\mathrm{gas} = 3.0 \times 10^5\,M_\odot$, and $m_\star = 6 \times 10^4\,M_\odot$, while the softening length is $\epsilon_\mathrm{DM}=860$ pc and $\epsilon_\mathrm{gas}=\epsilon_\star = 488$ pc. 

\begin{figure*}
\centering
  \includegraphics[width=15cm]{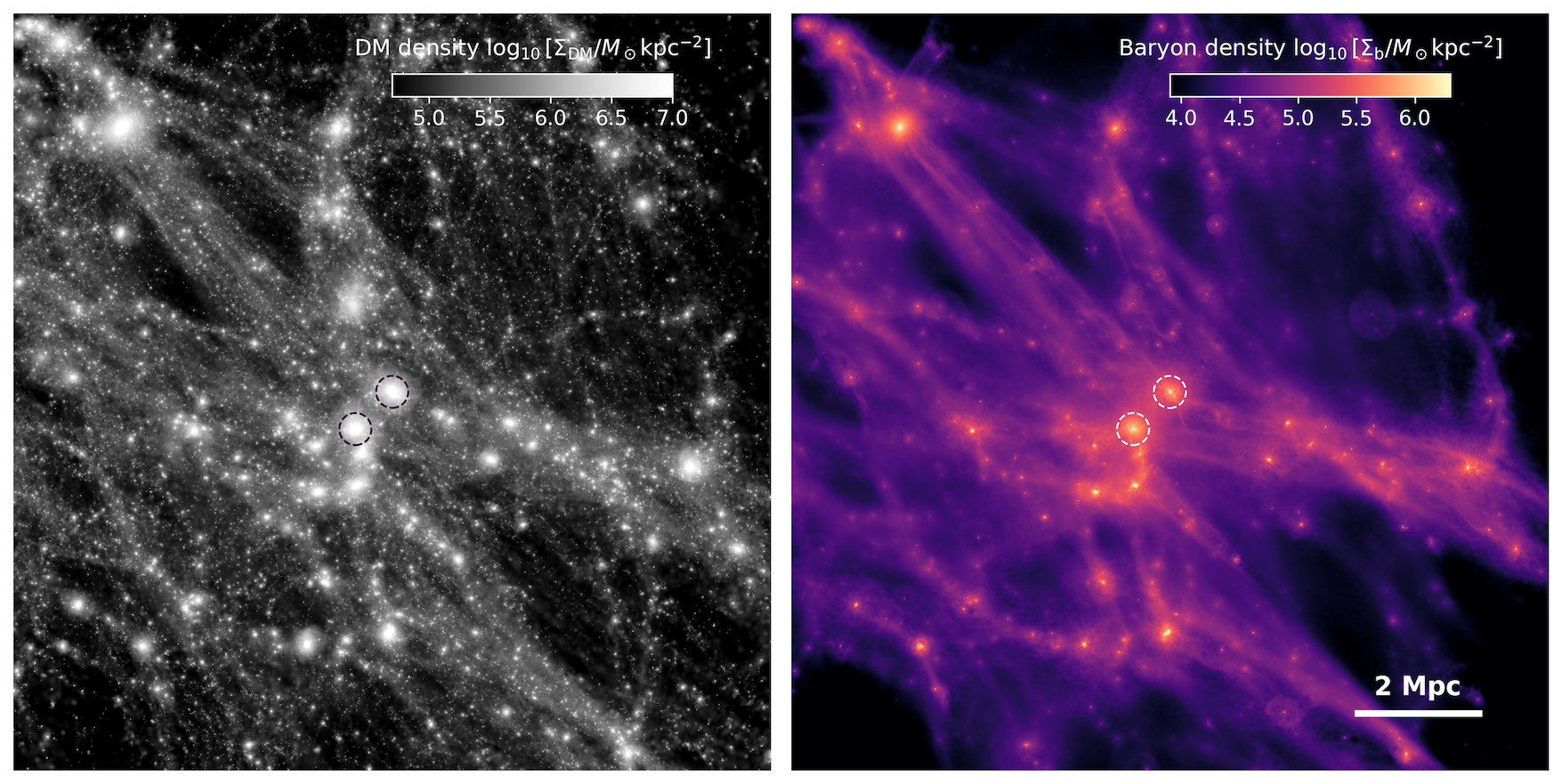}
  \caption{Projected mass distribution in the NIVARIA Local Group simulation. Left: dark matter surface density. Right: baryonic surface density (gas + stars). The maps show line-of-sight projections of the full simulation (high-resolution region only), centered on the Milky Way analogue. Both the Milky Way and M31 analogues are indicated with dashed circles. The Local Group environment is visible within the surrounding filamentary structure of the cosmic web.}
  \label{fig:LSS}
\end{figure*}

The initial conditions (ICs) are taken from the \texttt{HESTIA} project \citep{Libeskind2020}, and are constructed to replicate the main large-scale structures of the Local Volume at $z = 0$, including a Milky Way--Andromeda-like pair and a Virgo cluster-like halo, within a $\Lambda$CDM framework. In the following subsections we will briefly explain the specifics of the \texttt{HESTIA} project and the IC generation process, as well as the \texttt{Gasoline2} code used for running the simulation and the post-processing products. An overview of the resulting mass distribution in the high-resolution region is shown in \Fig{fig:LSS}.

\subsection{Initial conditions: The HESTIA project} \label{sec:simulations:hestia}

\texttt{HESTIA}, High-resolution Environmental Simulations of The Immediate Area, is the successor of the original CLUES project\footnote{https://www.clues-project.org/} (Constrained Local UniversE Simulations; \citealp{Gottloeber2010,Libeskind2010,Carlesi2016,Sorce2016}). The main goal of this project is to provide constrained simulations of the Local Universe, so that they can be used as a numerical laboratory for testing the current theoretical framework. 
We briefly explain here the main aspects of this project and how it connects to our \texttt{NIVARIA-LG} simulations. For further detail we refer the reader to the presentation paper by \citet{Libeskind2020}.

The ICs are generated to be consistent with the $\Lambda$CDM model and constrained by observational data of the local Universe. In this paradigm, the early Universe is characterized by fluctuations in density and peculiar velocity fields that can be modelled as Gaussian random fields \citep{Zeldovich1970,Bardeen1986}. At later times, gravitational instability drives the growth of structure, leading to a non-linear density field. However, on scales of a few Mpc, the velocity field remains close to Gaussian. This property enables the use of the Wiener Filter (WF), a linear Bayesian estimator, together with constrained realizations \citep{Hoffman1991,Zaroubi1995} to reconstruct the large-scale structure and peculiar velocity field \citep{Zaroubi1999}. 

The peculiar velocities of galaxies in the local Universe are taken from the CosmicFlows-2 (CF2) catalogue \citep{Tully2013}, which consists of over 8000 direct measurements of galaxy distances and redshifts out to a distance of $\sim250$ Mpc. Observational biases such as the `Malmquist bias' and lognormal error distribution have been considered and corrected for following \citet{Sorce2015}. Using these corrected peculiar velocities, together with the WF and constrained realizations technique, the large-scale constrained density field is reconstructed and subsequently rescaled to the simulation box.

Small-scale fluctuations are then added as random realizations consistent with the $\Lambda$CDM power spectrum \citep{Carlesi2016}. Since this procedure does not necessarily produce a Local Group-like object, the \texttt{HESTIA} approach consists of running many low-resolution realizations and selecting those that best resemble the local Universe. The identification of suitable ICs is based on three robust features on large scales: the Virgo cluster, the local void, and the local filament. On small scales, a Local Group is required to form near the center of the box, defined as a pair of infalling halos with similar masses and consistent with the observed properties of the Milky Way and M31. The selected ICs are then regenerated at higher resolution. In the case of the intermediate-resolution simulations used in this work, the resolution is increased within a sphere of radius 7.4 Mpc centered on the Local Group.

The main product of the \texttt{HESTIA} project is a set of constrained simulations of the Local Group, three of which run at high resolution, reaching down to $m_\mathrm{DM} = 1.5 \times 10^5\,M_\odot$ and 220 pc in spatial resolution, with an effective number of particles in the zoom-in region corresponding to $8192^3$ in the full box; and 13 realisations at intermediate resolution, with $4096^3$ effective particles in the zoom region. For the \texttt{NIVARIA-LG} simulation we present in this work, we select the ICs used to generate one of the intermediate resolution runs. This way, our simulation spans a region of roughly 6 Mpc in radius uncontaminated by lower-resolution elements, with a Local Group pair at the center surrounded by field and satellite dwarf galaxies.

\subsection{The NIVARIA implementation: Gasoline2} \label{sec:simulations:nivaria}


The \texttt{NIVARIA-LG} simulations are performed with \texttt{Gasoline2}, a modern smoothed-particle hydrodynamics (SPH) code widely used in galaxy formation studies. The adopted cosmology is a flat $\Lambda$CDM with cosmological parameters from \citet{Planck2014}: Hubble constant $H_0=100$ $h$ km s$^{-1}$ Mpc$^{-1}$ with $h=0.671$; dark energy density $\Omega_\Lambda=0.682$, matter density $\Omega_\mathrm{M}=0.318$ with baryon density $\Omega_\mathrm{b}=0.049$, power spectrum normalization $\sigma_8=0.8344$, and power spectrum slope $n=0.9624$. 

\texttt{Gasoline2} \citep{Wadsley2017} is an updated version of the original \texttt{Gasoline} code \citep{Wadsley2004}, incorporating several improvements in both hydrodynamics and subgrid physics, and is publicly available\footnote{\url{https://gasoline-code.com}}. \texttt{Gasoline} has been built upon the \texttt{pkdgrav} parallel tree code \citep{Stadel2001}, and retains its core architecture, including the Kick-Drift-Kick (KDK) integration scheme, gravity solver and tree structure. 
The code employs an advanced SPH formulation that addresses well-known shortcomings of traditional SPH methods, including improved treatment of fluid mixing, contact discontinuities, and instabilities. In particular, it adopts a geometric-density-average force expression, a Wendland kernel to suppress particle pairing, and time-dependent artificial viscosity and thermal diffusion terms that enhance the capture of shocks and mixing processes.

Cooling is modelled through hydrogen, helium, and metal-line processes in the presence of a spatially uniform UV ionizing background, following the prescriptions of \citet{Shen2010}. The cooling rates are computed using \texttt{CLOUDY} \citep{Ferland1998}, incorporating photoionisation and heating from the \citet{HaardtMadau2012} UV background, and Compton cooling. This UV background leads to the reionisation of hydrogen in the intergalactic medium by $z\sim6$. 

The implementation of star formation and feedback follows the model used in the \texttt{NIHAO} \citep[Numerical Investigations of Hundred Astrophysical Objects;][]{Wang2015} and \texttt{MaGICC} \citep[Making Galaxies In a Cosmological Context;][]{Stinson2013} projects, which have been shown to successfully reproduce galaxy scaling relations over a wide mass range \citep{Brook2012,Maccio2012,Wang2015,Dutton2017}. Gas is allowed to form stars when it is cold and dense enough, following the Kennicutt-Schmidt relation. Specifically, gas must have a temperature $T < 1.5\times10^4$ K and exceed a density threshold number set at $n_\mathrm{th}=10$ cm$^{-3}$. Star formation then follows an initial mass function (IMF) from \citet{Chabrier2003}.


Stellar feedback is implemented in two main phases. Early stellar feedback comes from massive stars before they explode as supernovae \citep{Stinson2013}, injecting about 13 per cent of their total stellar energy into the surrounding gas while leaving radiative cooling on. The second phase is supernova feedback, which occurs when stars of $8-40$ $M_\odot$ explode, releasing both energy and metals into the interstellar medium. This feedback is implemented using a thermal blast-wave approach \citep{Stinson2006}. To prevent dense gas from radiating away this energy too quickly, cooling is temporarily delayed ($\sim$30 Myr) in the blast region.  Metals are ejected from Type Ia and Type II supernovae, and the stellar winds driven from asymptotic giant branch (AGB) stars, and distributed to the nearest gas particles using the smoothing kernel. 
This feedback implementation enables the development of bursty star formation and strong galactic outflows, processes that can induce the formation of dark matter cores in dwarf galaxies at a specific mass range, thus helping alleviate some small-scale tensions within the $\Lambda$CDM paradigm \citep[e.g.][]{DiCintio2014a,BrookDiCintio2015}.

We further note here that the NIVARIA project includes the study of AGN feedback in dwarfs. Therefore, we use a novel version of \texttt{Gasoline2} that allows the inclusion of BHs and AGN feedback \citep{Blank2019,Waterval2024}. However, for this initial run we have deactivated that option, and so our galaxies do not contain BHs. In a forthcoming paper we will present the results of running the same simulation with the BH module enabled, so that the effects of AGN feedback in Local Group dwarfs can be  tested.

\subsection{Halo catalogues and merger trees} \label{sec:simulations:ahf}

Dark matter halos in the simulation are identified using the Amiga Halo Finder (AHF;\footnote{\url{https://popia.ft.uam.es/AHF}} \citealp{Gill2014,KnollmannKnebe2009}). This algorithm detects local overdensities within the simulation volume using a procedure similar to adaptive mesh refinement (AMR), recursively refining regions where the total matter density exceeds a given threshold. Once these regions are identified, particles of all components (dark matter, gas, stars, and black holes) are assigned to gravitationally bound structures.
Halos are defined as spherical regions containing at least 20 particles and a mean density of $\Delta \rho_{\mathrm{crit}}(z)$, where $\rho_{\mathrm{crit}}$ is the critical density of the Universe. $\Delta$ is the overdensity parameter, for which we adopt a fixed value of $\Delta = 200$. This defines the corresponding halo mass and radius, $M_{200}$ and $R_{200}$.
Subhalos are defined as smaller bound systems located within the radius of a more massive host halo, and their mass is computed as the total gravitationally bound mass associated with their centers. 

To trace the evolution of the identified halos through the different snapshots, we use merger trees created with the \textsc{MergerTree} tool, that comes with the AHF package. \textsc{MergerTree} follows each halo identified at $z=0$ backwards in time, identifying its progenitors in previous snapshots based on a merit function. It also allows for skipping snapshots if no progenitor is found in the immedate previous snapshot. 

\section{Galaxy properties} \label{sec:galaxies}

In this section we first examine the global scaling relations of the galaxy population in the \texttt{NIVARIA-LG} simulations as a consistency check of the model. We then define and compute the main structural and kinematic quantities used throughout the paper.

From the full galaxy sample, we select systems within the high-resolution region of the Local Group, requiring that they are not contaminated by low-resolution particles\footnote{We select galaxies with a fraction of high-resolution particles larger than 0.99}. We further restrict the sample to isolated galaxies, excluding satellites of more massive systems, in order to focus on objects whose properties are not significantly affected by interactions with a massive host. 
This choice also facilitates comparison with observational samples that are predominantly composed of field galaxies.

\subsection{Global scaling relations}


\begin{figure*}
\centering
  \includegraphics[width=18cm]{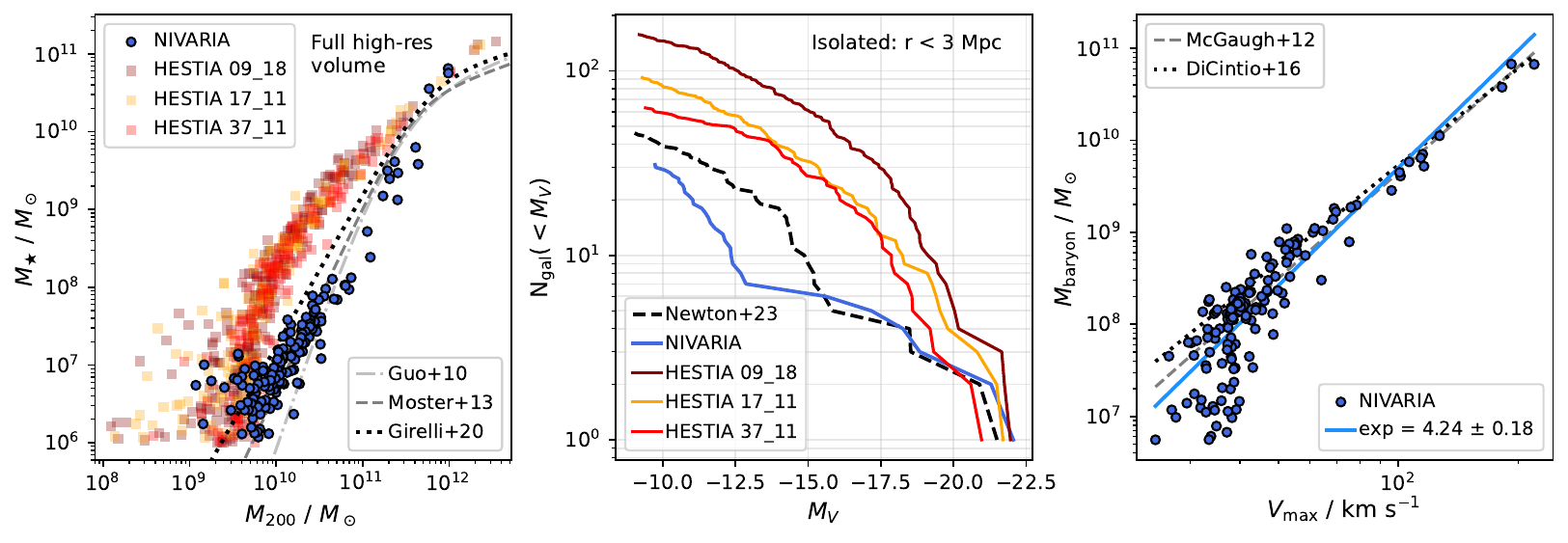}
  \caption{Left, Stellar-to-halo mass relation for isolated galaxies. \texttt{NIVARIA-LG} galaxies are shown as blue dots, while galaxies from the three high-resolution \texttt{HESTIA} simulations are depicted as squares. Abundance matching relations from \citet{Guo2010}, \citet{Moster2013}, and \citet{Girelli2020} are shown as grey dash-dotted, dashed, and dotted lines, respectively. Middle, luminosity function of the \texttt{NIVARIA-LG} and \texttt{HESTIA} simulations (solid lines). Dashed line is the observed luminosity function computed from the compilation by \citet{Newton2023}. Right, Baryonic Tully-Fisher relation of the simulated \texttt{NIVARIA-LG} galaxies. Baryonic mass is computed as the sum of HI gas and stellar mass, while for the velocity we use the maximum value of the rotation curve. Solid blue line indicates the fit of the \texttt{NIVARIA-LG} data to a power law, while gray dashed and dotted lines correspond to reference values from observations \citep{McGaugh2012} and semi-analytic models \citep{DiCintioLelli2016}.}
  \label{fig:mstar-mhalo}
\end{figure*}

We first present the stellar-to-halo mass relation of the \texttt{NIVARIA-LG} simulations in the left panel of \Fig{fig:mstar-mhalo}. The blue dots represent all galaxies in \texttt{NIVARIA-LG}. For reference, abundance matching relations from \citet{Guo2010}, \citet{Moster2013} and \citet{Girelli2020} are shown as dash-dotted, dashed and dotted lines, respectively. We find that \texttt{NIVARIA-LG} galaxies broadly follow these relations over the full mass range considered, indicating that the simulations reproduce the observed galaxy formation efficiency.
For comparison, we overplot the same relation for galaxies from the three high-resolution \texttt{HESTIA} simulations, including only galaxies in the same stellar mass range as \texttt{NIVARIA-LG}. In this case, we find a systematic overproduction of stars, particularly for halos with masses between $10^{10}$ and $10^{11}$ $M_\odot$. The improved agreement in \texttt{NIVARIA-LG} suggests that the adopted subgrid physics leads to a more realistic regulation of star formation in low-mass halos. 

In the second panel of \Fig{fig:mstar-mhalo} we show the cumulative $V$-band luminosity function of isolated galaxies, considering only systems located within 3 Mpc of the MW analogue. The blue curve corresponds to \texttt{NIVARIA-LG}, while the dark red, orange, and red curves show the results from the three high-resolution \texttt{HESTIA} simulations. As a reference, the black dashed line shows the observed luminosity function from the compilation of \citet{Newton2023}, combining galaxies from \citet{McConnachie2012} and \citet{Savino2022}. We find that the \texttt{NIVARIA-LG} luminosity function is in good agreement with the observational estimate over most of the magnitude range probed, reproducing both the abundance of bright galaxies and the relatively shallow increase towards faint magnitudes. In contrast, all three \texttt{HESTIA} realizations predict substantially larger numbers of galaxies, particularly at the faint end ($M_V \gtrsim -18$), with excess counts reaching more than an order of magnitude for the faintest systems. 


It is important to note that the total mass of the simulated Local Group environment directly influences the overall abundance of galaxies therein. The total enclosed mass (dark matter, gas and stars) within 3 Mpc of the MW analogue, for the three \texttt{HESTIA} realisations, is of  $8.9 \times 10^{12}$, $5.8 \times 10^{12}$, and $3.4 \times 10^{12}$ $M_\odot$, respectively, whereas within the \texttt{NIVARIA-LG} environment is  of $2.0 \times 10^{12}$ $M_\odot$. Therefore, the higher number of galaxies in the \texttt{HESTIA} runs is partially driven by their more massive environments. Nevertheless, this overabundance is also fundamentally tied to the enhanced stellar masses seen in the left panel, indicating that \texttt{NIVARIA-LG} provides a more realistic suppression of star formation in low-mass halos, resulting in a dwarf galaxy population that better matches current observations.


To further assess whether our simulated galaxies provide a reliable representation of observed systems, we examine their Baryonic Tully–Fisher relation (BTFR) in the right panel of \Fig{fig:mstar-mhalo}, which relates the total baryonic mass to the characteristic rotation velocity of galaxies \citep{McGaugh2012}. The baryonic mass is computed as the sum of the stellar mass and neutral hydrogen content. For the rotation velocity, we adopt the maximum value of the rotation curve, $V_\mathrm{max}$, as a robust measure of the galaxy's potential well (see \Sec{sec:methods:RCs} for the full methodology). While this choice differs from typical observational definitions at a fixed radius, it allows for a consistent comparison across galaxies of different masses.

The right panel of \Fig{fig:mstar-mhalo} shows the resulting BTFR for \texttt{NIVARIA-LG} galaxies (blue dots), compared to observations from \citet{McGaugh2012} and the semi-analytic model of \citet{DiCintioLelli2016}, depicted as dashed and dotted lines, respectively. Fitting a power-law to the \texttt{NIVARIA-LG} galaxies, we obtain a slope of $4.24 \pm 0.18$, steeper than the observed values of 3.82 and 3.50 reported by \citet{McGaugh2012} and \citet{DiCintioLelli2016}. While most of the galaxies match the depicted relations closely, specially at the high-mass end, the scatter increases at the low-mass end. Some low-mass systems fall below the expected velocities, thus increasing the fitted slope. This steepening of the slope can be partly due to the chosen $V_\mathrm{max}$ definition. Observed dwarf galaxy rotation curves often do not extend to the radius where the halo reaches it maximum circular velocity, and so $V_\mathrm{max}$ is larger in the simulated galaxies \citep{Brook2016a,Brook2016b}. Despite this, the overall normalisation and trend of the BTFR in \texttt{NIVARIA-LG} remain broadly consistent with observations. 


\subsection{Galaxy structure and kinematics}
For the galaxy sample presented above, we compute a set of structural and kinematic properties that allow us to characterise the diversity of galaxy populations in the simulations. In the following subsections we describe the methods used to derive the effective radius, rotation curves, and dark matter density profiles. 

\subsubsection{Effective radius} \label{sec:methods:sizes}

There are several ways to define the size of a galaxy (e.g. based on stellar mass, light profiles, or kinematics). We here employ the effective radius, $r_\mathrm{eff}$, defined as the projected radius enclosing half of the total luminosity of the galaxy. 
To compute $r_\mathrm{eff}$, we first rotate each galaxy into a face-on configuration by aligning the stellar angular momentum with the line of sight (the $z$-axis). We then calculate the luminosity in the $r$-band (ignoring dust attenuation), and determine the 2D radius enclosing half of the total emitted light.

\begin{figure}
\centering
  \includegraphics[width=9cm]{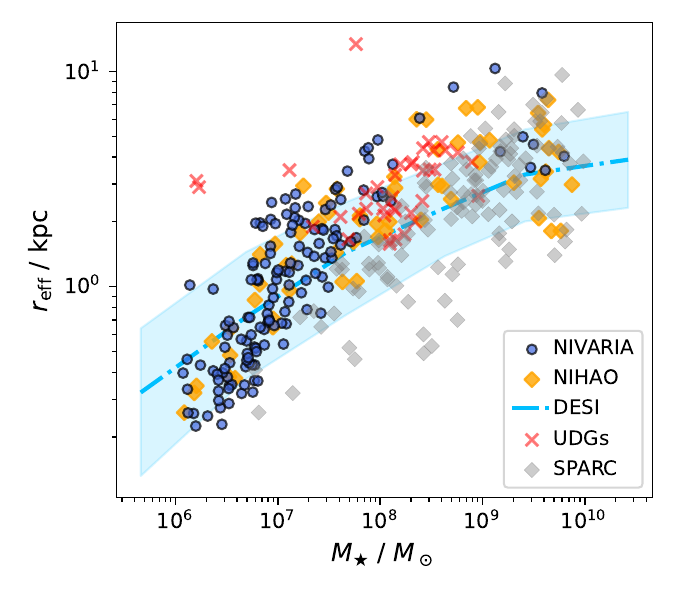}
  \vspace*{-0.6cm}
  \caption{Effective radius, $r_\mathrm{eff}$, as a function of stellar mass. \texttt{NIVARIA-LG} galaxies with $M_\star<10^{10}\,M_\odot$ are shown as blue dots, while orange diamonds indicate results from the original \texttt{NIHAO} run. The cyan line and shaded region show the median and 16th-84th percentile range of the DESI DR1 sample selected by \citet{Benavides2026}. Red crosses and gray diamonds correspond to observed UDGs compiled by \citet{Gannon2024} and SPARC galaxies \citep{Lelli2016}, respectively.}
  \label{fig:size-mstar}
\end{figure}

\Fig{fig:size-mstar} shows the relation between the size and the stellar mass of the galaxies in the \texttt{NIVARIA-LG} simulations (blue dots). We focus on the regime of dwarf galaxies, with $M_\star<10^{10} M_\odot$. At low and intermediate masses, a clear positive correlation can be seen between size and mass, while at higher masses ($M_\star \gtrsim 10^8 M_\odot$) the slope of the correlation is reduced and the scatter increased. 

For comparison, we include galaxy sizes computed following the same procedure for the original \texttt{NIHAO} simulations, depicted as orange diamonds. Our galaxies are broadly consistent with \texttt{NIHAO} ones, although \texttt{NIVARIA-LG} exhibits increased scatter at intermediate masses. Since the code employed for both simulations is the same, \texttt{Gasoline2}, the origin of the increased scatter is likely related to factors other than the choice of code itself. 
As a measure of the scatter in the size-mass relation in the \texttt{NIVARIA-LG} simulations, in \Tab{tab:mass_radius_bins} we separate the \texttt{NIVARIA-LG} galaxies in bins of stellar mass, and indicate the maximum and minimum value of $r_\mathrm{eff}$ present within each bin. This table shows how, at a fixed stellar mass, the size of galaxies can vary of an order of magnitude.

We compare our results in \Fig{fig:size-mstar} with several observational data sets in the same stellar mass range. The cyan line shows the data from DESI DR1 \citep{DESI2026}, selected by \citet{Benavides2026}, which includes more than $600\,000$ galaxies, with the shaded region indicating the 16th-84th percentile range of the data. Gray diamonds correspond to SPARC galaxies \citep{Lelli2016}, for which stellar mass is derived assuming a constant stellar mass-to-light ratio of 0.5. We also include data of ultra-diffuse galaxies (UDGs), compiled by \citet{Gannon2024}. We see that, for low masses, $M_\star\lesssim10^8 M_\odot$, where the \texttt{NIVARIA-LG} sample is better populated, the simulated galaxies are broadly consistent with the DESI sample. SPARC galaxies occupy a region of smaller sizes, although this sample may be biased towards brighter galaxies by construction. This difference is more apparent when comparing to the UDG sample, which lies closer to the \texttt{NIVARIA-LG} and \texttt{NIHAO} galaxies, suggesting that these simulations reproduce this population of extended diffuse systems relatively well. 
Unlike previous simulation suites, our latest \texttt{NIVARIA-LG} simulations successfully reproduce, at fixed stellar mass, \textit{the diversity of dwarf galaxies morphologies}, captured by their  broad range of galaxy sizes, from compact to diffuse systems, and being broadly consistent with state-of-the-art observations.

\begin{table}
\caption{Minimum and maximum $r_\mathrm{eff}$ per stellar mass bin in isolated dwarf galaxies from \texttt{NIVARIA-LG} simulations.}
\label{tab:mass_radius_bins}
\centering
\small
\begin{tabular}{ccc}
\hline
$\log(M_\star/M_\odot)$ & $r_{\rm eff,min}$ (kpc) & $r_{\rm eff,max}$ (kpc)             \\
\hline
6--7   & 0.22 & 2.64 \\
7--8   & 0.54 & 5.16 \\
8--9   & 2.49 & 8.46 \\
9--10  & 3.47 & 10.32 \\
\hline
\end{tabular}
\end{table}

\subsubsection{Rotation curves} \label{sec:methods:RCs}
In this subsection we describe how the galaxy rotation curves are computed. We employ the enclosed-mass method, defining the circular velocity as:
\begin{equation}
    V_\mathrm{circ}(r) = \sqrt{\frac{GM(<r)}{r}},
\end{equation}
where $M(<r)$ is the total mass enclosed in a radius $r$ from the galaxy centre and $G$ is the gravitational constant. The centre is determined using the shrinking-sphere method of \citet{Power2003}.
The rotation curves are computed using 50 logarithmically spaced radial bins, extending out to 15 kpc for galaxies up to $M_\star = 10^9 M_\odot$ and 30 kpc for the most massive galaxies. These radial ranges ensure that the maximum of the rotation curve is reached for all galaxies in the sample.
Alternative approaches, such as directly computing the radial derivative of the gravitational potential in the disc midplane \citep[e.g.][]{Santos-Santos2018,Cruz2025}, were also considered. However, we find this method more sensitive to local particle noise and hence, the enclosed mass method better to handle more irregular galaxies and provide more stable rotation curves. 

Once the full rotation curves are computed, we evaluate them at $R_\mathrm{last}$, which we estimate from the stellar mass following Eq. 1 in \citet{Santos-Santos2018}. This empirical relation was obtained by fitting the outermost radius measured in observed HI rotation curves as a function of stellar mass and provides an estimate of the radius typically reached by HI observations for galaxies of similar mass. We additionally compute the circular velocity at a fixed radius of 2 kpc as a proxy for the inner rotation curve. Adopting a fixed radius allows for a homogeneous comparison across galaxies and ensures that the measurement lies within a well-resolved region for all systems considered. 
The impact of these choices is explored in Appendix~\ref{appendix:RCs}, where we compare different methods for calculating the rotation curves and alternative definitions of the characteristic radii and velocities.


\subsubsection{Dark matter density profiles} \label{sec:methods:DMprofiles}

We compute dark matter density profiles to quantify the inner structure of the halos and assess variations in their central density distribution. The profiles are constructed using logarithmically spaced spherical radial bins. The inner structure of the halo is quantified by measuring the local logarithmic slope of the density profile in the radial ranges 1–2$\%$ and 2–3$\%$ of the virial radius, which we denote as $\alpha_{12}$ and $\alpha_{23}$, respectively. These ranges are chosen to probe the central region of the halo while remaining above the resolution limit.
This first run of the \texttt{NIVARIA-LG} simulations, using a very similar 
 feedback scheme as the original \texttt{NIHAO} simulations, produces galaxies 
whose dark matter profiles have a cored distribution in the same stellar mass 
range, i.e. for galaxies with stellar masses between $10^6$ and $10^{10}$ \Msun, 
as shown in \citet{Tollet2016}.

\begin{figure*}
\centering
  \includegraphics[width=15cm]{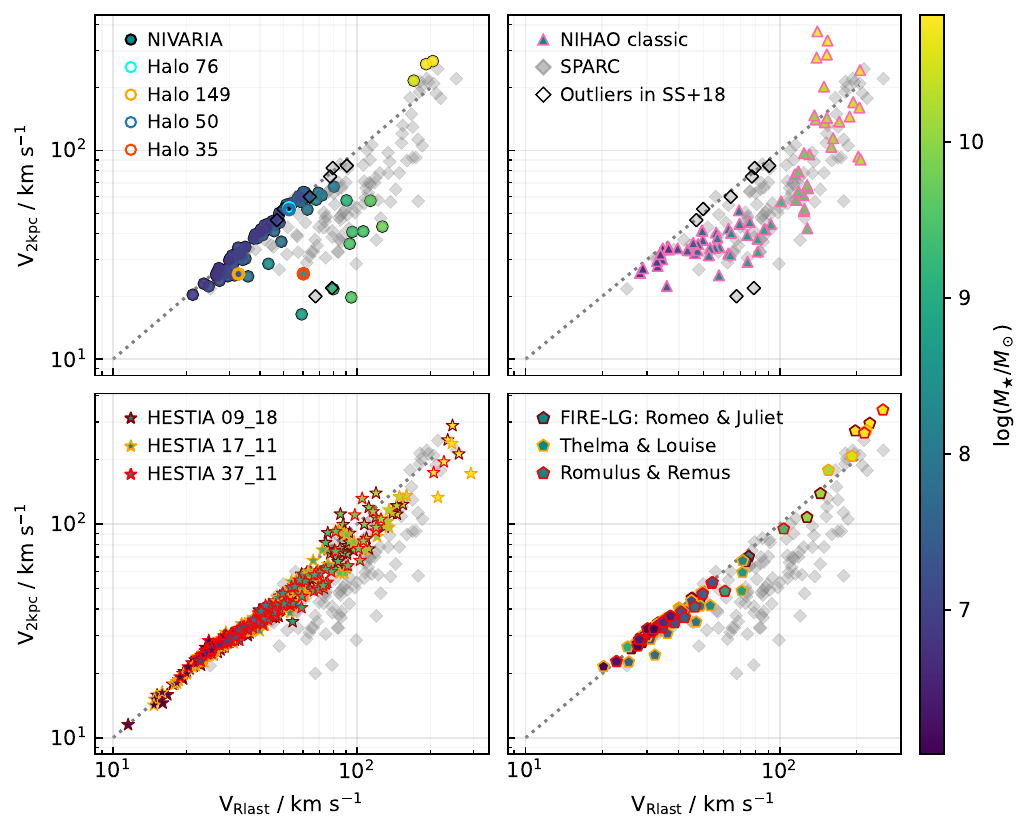}
  \caption{$V_\mathrm{2kpc}-V_\mathrm{Rlast}$ relation for the \texttt{NIVARIA-LG} simulated galaxy sample (circles) and SPARC observations (gray diamonds). Top left includes  \texttt{NIVARIA-LG} and SPARC galaxies, highlighting the halos featured in \Fig{fig:example-bin2}; top right includes  \texttt{NIHAO} galaxies for comparison, depicted as triangles with pink edges. Bordered diamonds are those marked as outliers by \citet{Santos-Santos2018} when comparing to the \texttt{NIHAO} sample. All simulated galaxies are coloured by their stellar mass. Bottom row shows the same comparison but against the three high resolution \texttt{HESTIA} simulations on the left and the three FIRE Local Groups on the right.}
  \label{fig:RCs-comparison}
\end{figure*}

\section{Results} \label{sec:results}

In this section we present our results regarding the rotation curves of the galaxies in the \texttt{NIVARIA-LG} simulations. We first present our results and compare to other simulations. We then discuss the results in their physical context, trying to understand them and connecting them to the underlying physical scenario. 

For the full analysis, we restrict our sample to galaxies with a number of stellar particles $N_\star \geq 100$, which leaves us with 104 galaxies with stellar masses between  $\sim 6 \times 10^6$ and $7\times10^{10}\,M_\odot$.

\subsection{Diversity of rotation curves} \label{sec:results:RCs}

Following the procedure described in \Sec{sec:methods:RCs}, we compute full rotation curves for all galaxies in the \texttt{NIVARIA-LG} sample, as well as the characteristic velocities $V_\mathrm{2kpc}$ and $V_\mathrm{Rlast}$. In the top-left panel of \Fig{fig:RCs-comparison} we show the resulting distribution, coloured by stellar mass. 
We compare our results to observational data from the SPARC sample \citep{Lelli2016}, which consists of 175 nearby galaxies with high-quality rotation curves from HI and/or H$\alpha$ studies. From this sample, we select galaxies with inclination $i>30^{\circ}$ and quality flag $Q<3$, following previous studies \citep{Santos-Santos2020,Zentner2022,Cruz2025}, and restrict the sample to galaxies with stellar masses within the range spanned by the \texttt{NIVARIA-LG} sample. We compute $V_\mathrm{2kpc}$ and $V_\mathrm{Rlast}$ for the observed rotation curves following the same procedure used for \texttt{NIVARIA-LG} galaxies.

Overall, the \texttt{NIVARIA-LG} galaxies span a broad range of rotation curve behaviours, from rapidly rising systems that lie close to the 1:1 relation to more gradually rising curves populating the lower part of the diagram. While the sample does not fully cover the highest-mass regime due to the limited volume of Local Group environments, it captures much of the diversity seen in SPARC galaxies.

We next compare \texttt{NIVARIA-LG} to other simulation suites. In the top-right panel of \Fig{fig:RCs-comparison}, we show the \texttt{NIHAO} sample (pink-edged triangles), analysed using the same pipeline as \texttt{NIVARIA}. Both simulations are based on the same hydrodynamical code, \texttt{Gasoline2}, but differ in their selection criteria and cosmological environment. \texttt{NIHAO} consists of isolated central galaxies drawn from cosmological volumes, while \texttt{NIVARIA-LG} is a constrained simulation of a Local Group-like environment. Additional differences in resolution and baryonic implementation are also present. 
We also highlight the subset of SPARC galaxies identified as outliers with respect to \texttt{NIHAO} by \citet{Santos-Santos2018}. These systems fall within the range populated by \texttt{NIVARIA-LG} galaxies, indicating improved agreement for this data set. 

In the bottom row of \Fig{fig:RCs-comparison}, we extend the comparison to Local Group analogues from the \texttt{HESTIA} and FIRE simulations. The \texttt{HESTIA} suite \citep{Libeskind2020}, as \texttt{NIVARIA-LG}, also uses constrained Local Group initial conditions, but employs a different hydrodynamical code (AREPO, \citealp{Springel2010}) and galaxy formation model (Auriga, \citealp{Grand2017}). Following the same methodology, we compute $V_\mathrm{2kpc}$ and $V_\mathrm{Rlast}$ for the three high resolution \texttt{HESTIA} simulations. 
FIRE-LG simulations \citep{Garrison-Kimmel2019} are not constrained simulations but rather consist of Local Group-like galaxy pairs selected from a larger cosmological volume and resimulated at higher resolution.  

Both \texttt{HESTIA} and FIRE-LG simulations preferentially populate the region corresponding to rapidly rising rotation curves, with no systems occupying the low $V_\mathrm{2kpc}$ regime. 
This difference with \texttt{NIVARIA-LG} may reflect a combination of modelling choices, including differences in the implemented galaxy formation models and feedback prescriptions. Previous studies have shown that feedback can significantly affect the inner structure of galaxies and the resulting rotation curve shapes \citep{DiCintio2014b,DiCintio2014a, Tollet2016}.

Overall, the comparison across simulation suites highlights that the distribution of rotation curve shapes is sensitive to a combination of modelling choices, numerical methods, and environmental selection effects.

\subsection{Illustrative examples of the diversity} \label{sec:results:examples}

\begin{figure*}

\textbf{a) Different $r_\mathrm{eff}$, different rotation curve shape:}

{\centering
\includegraphics[width=17.5cm]{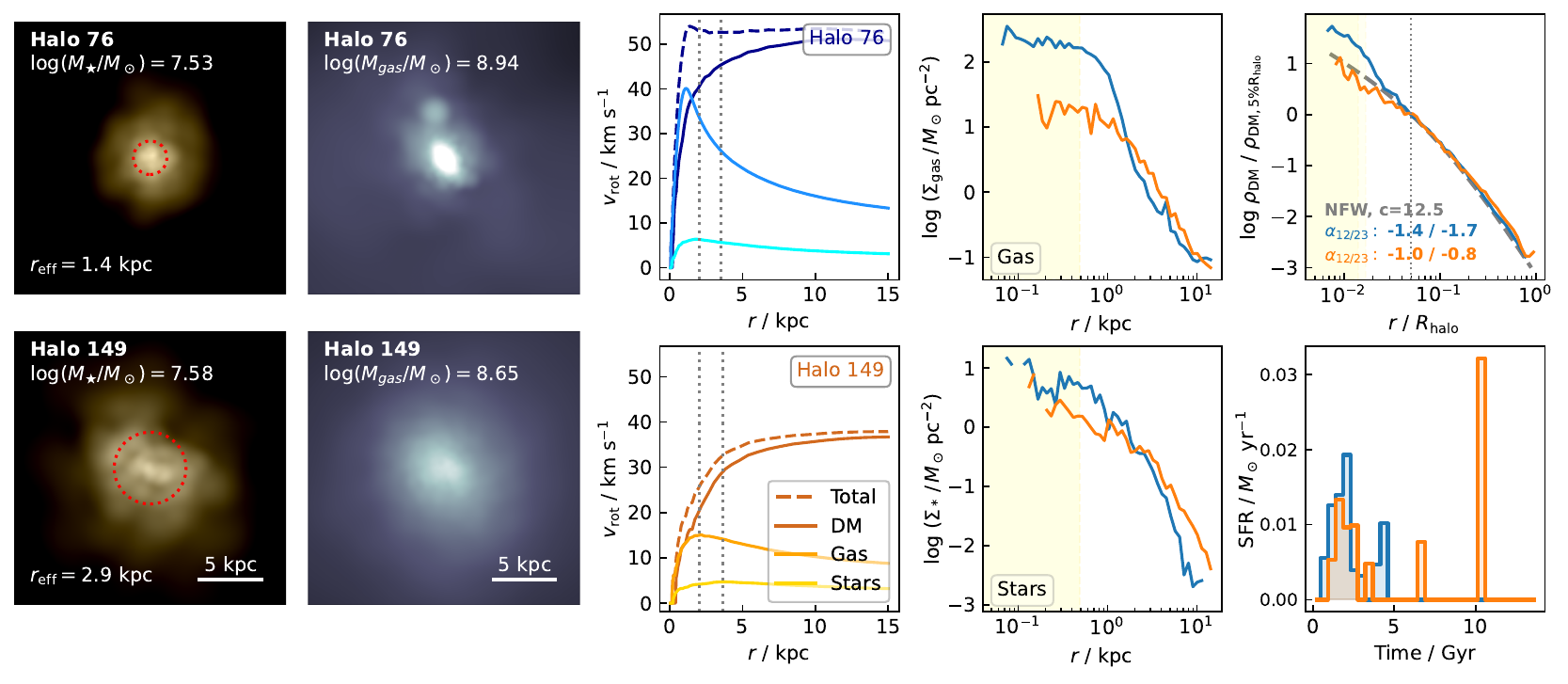}
}

\vspace*{-0.2cm}
\textbf{b) Similar $r_\mathrm{eff}$, different rotation curve shape:}

{\centering
\includegraphics[width=17.5cm]{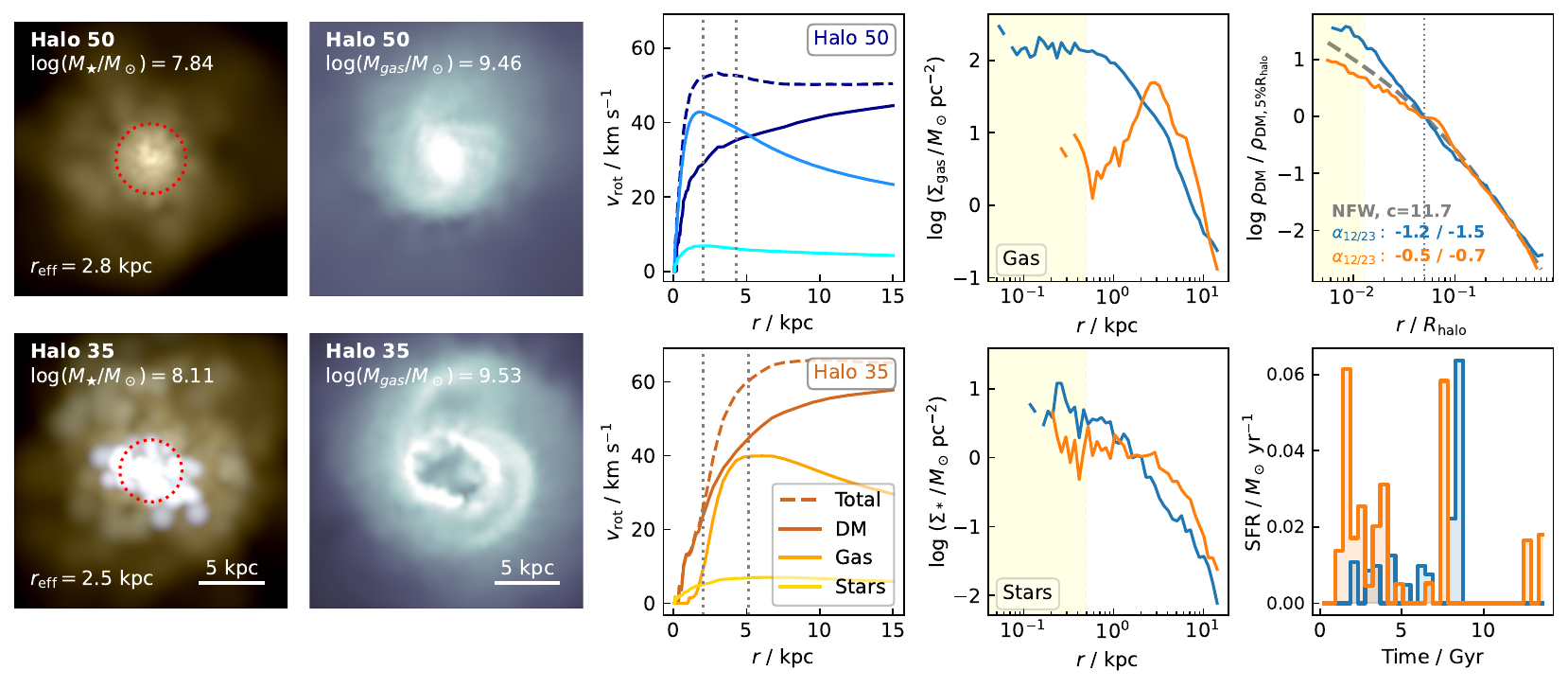}
}
\caption{Comparison of two pairs of galaxies as illustrative examples of the diversity of rotation curves. On top, halos 76 (blue) and 149 (orange), with very similar stellar masses but different effective radii, giving rise to different rotation curve shape. On the bottom, halos 50 (blue) and 35 (orange), which instead have a very similar effective radius. The structure of the figure is the same for both comparisons. First column: face-on rendering of the stellar light distribution; red dotted circles mark the effective radius, $r_\mathrm{eff}$. Second column: face-on gas density maps. Total stellar and gas mass are indicated in the corresponding panels. Middle column: rotation curve of the galaxies, decomposed into total (dashed), dark matter, gas, and stellar contributions; vertical lines denote $r = 2$ kpc and $R_\mathrm{last}$. Fourth column: radial profiles of gas (top) and stellar (bottom) surface density, with shaded regions indicating the softening $\epsilon_\mathrm{gas}=488$ pc. Last column (top): Logarithmic dark-matter density profiles, normalised to their value at $0.05R_\mathrm{halo}$; the quoted values correspond to inner slopes measured over the radial ranges $1-2\%R_\mathrm{halo}$ and $2 - 3\%R_\mathrm{halo}$. The dashed grey line depicts a NFW profile for a halo with similar mass ($M_\mathrm{halo} = 10^{10.20}M_\odot$ for halos 76-149, and $10^{10.55}M_\odot$ for halos 50-35) and concentration $c$, derived following \citet{DuttonMaccio2014}. Last column (bottom): Star formation histories of the two galaxies.}
\label{fig:example-bin2}
\end{figure*}

To illustrate the diversity of rotation curves present in the \texttt{NIVARIA-LG} simulations, \Fig{fig:example-bin2} shows two pairs of galaxies selected to have comparable stellar masses but different shape of their rotation curve. For each of the panels a) and b), the structure of the figure is the same: first column shows a face-on rendering of the stellar light distribution, with the effective radius depicted as a red dotted circle, while the second column shows face-on gas density maps. The third column shows the rotation curves for the galaxies (see \Sec{sec:methods:RCs}), separated into total (dashed lines) and different components (solid lines), as indicated. Vertical lines indicate the radii corresponding to 2 kpc and $R_\mathrm{last}$ for the two galaxies, which probe different regimes of the rotation curve and vary systematically between galaxies. The fourth column shows the gas (top) and stellar (bottom) surface density radial profiles, each coloured line depicting the corresponding halo (76 in blue, 149 in orange). Similarly, the last column (top panel) shows the dark matter density profiles, normalised to their value at 5$\%$ of $R_{200}$ for ease of comparison. The corresponding inner slopes are indicated in the figure. For reference, the dashed gray line depicts the NFW profile for a halo with halo mass and radius corresponding to the mean of both halos, and concentration computed following the relation from \citet{DuttonMaccio2014}. The bottom panel in the last column shows the star formation histories of the galaxies.

The first comparison (top panel, a), halos 76 and 149) consists of galaxies with similar stellar masses, ($\log M_\star \sim 7.5$) but different effective radii. Despite their comparable stellar masses, halo 149 is approximately twice as extended as halo 76 and exhibits a significantly more slowly rising rotation curve. The gas component, while more massive in halo 76, is also more extended and diffuse in halo 149. The inner density profile, while similar for both galaxies, is slightly shallower for halo 149. This example highlights the connection between the spatial distribution of baryons and the shape of galaxy rotation curves.

The second comparison (bottom panel, b), halos 50 and 35) presents a different manifestation of the diversity. In this case, the two galaxies have comparable stellar masses and similar effective radii, but their rotation curves remain substantially different. Halo 35 exhibits a slowly rising rotation curve, while halo 50 reaches its maximum velocity at much smaller radii. The difference between these systems is reflected in their gas distributions: halo 35 shows a reduced central gas density, consistent with a cavity-like structure, while halo 50 retains a more centrally concentrated gas distribution. These differences can be associated with distinct recent star formation activity, as illustrated in the star formation histories and stellar light distributions. Moreover, the inner region of the dark matter density profiles also differs for both galaxies, with halo 35 presenting a shallower slope. 
In general, the more extended the star formation history, the more extended 
the gas distribution, the more cored the dark matter halo, and the shallower 
the inner slope, in agreement with several previous studies \citep{DiCintio2017,Muni2025,Sarrato-Alos2026}, based on 
the understanding that baryon feedback-induced cores are created through 
repeated gas outflows that effectively modify the potential well at the 
centre of galaxies \citep{PontzenGovernato2014}.

In both comparisons, the differences in rotation curve shape track systematic variations in the underlying mass distribution. Galaxies with more slowly rising rotation curves (halos 149 and 35) seem to exhibit more extended gas configurations and reduced central dark matter concentrations, despite having similar global stellar properties. This points to changes in the internal baryonic and dark matter structure as the main drivers of the observed diversity, and to the fact  that galaxy size alone cannot  account for the observed diversity in rotation curve shapes.

These examples indicate that rotation curve shapes are not determined by any single global galaxy property, but instead reflect combined variations in several structural components. In the following section, we quantify these relationships across the full \texttt{NIVARIA-LG} sample to identify the dominant correlations with rotation curve diversity.

\subsection{Statistical study of the diversity} \label{sec:results:structural}

\begin{figure*}
\centering
  \includegraphics[width=16cm]{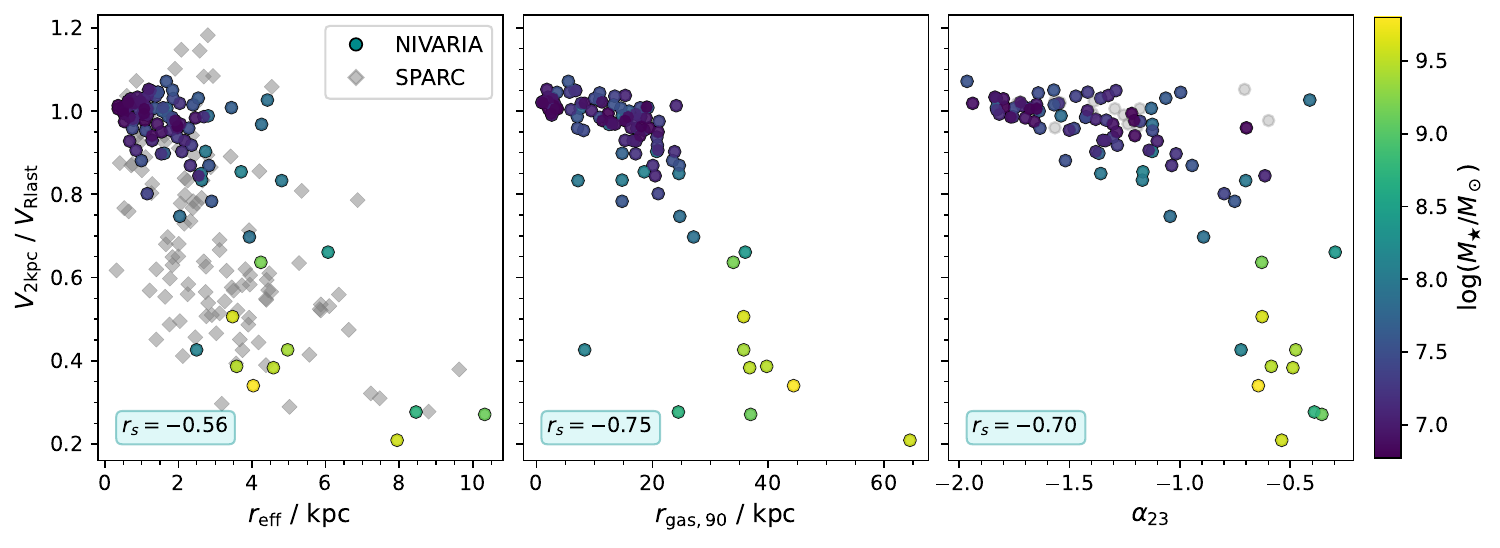}
  \vspace*{-0.2cm}
  \caption{Left, relation between the ratio $V_\mathrm{2kpc}/V_\mathrm{Rlast}$, as a measure of the shape of the rotation curve, and the effective radius of the galaxy, $r_\mathrm{eff}$. Circles are for the \texttt{NIVARIA-LG} galaxies, while gray diamonds show observations from SPARC. Center, relation between $V_\mathrm{2kpc}/V_\mathrm{Rlast}$ and the radius that encloses 90$\%$ of the total gas mass, $r_\mathrm{gas,90}$, for \texttt{NIVARIA-LG} galaxies. Right, relation between  $V_\mathrm{2kpc}/V_\mathrm{Rlast}$ and the slope of the dark matter density profile between 2 and 3 per cent of the halo radius, $\alpha_{23}$. Light gray points are galaxies for which $2\%$ of $R_\mathrm{200}$ is smaller than the softening of the simulation $\epsilon_\mathrm{DM}$. In all panels, circles are coloured by stellar mass. Spearman correlation coefficients between the two corresponding variables are indicated in the bottom of each panel.}
  \label{fig:RCcoef-reff}
\end{figure*}

In this section we examine the structural properties associated with the diversity of galaxy rotation curves in the \texttt{NIVARIA-LG} simulations. We quantify the shape of the rotation curve using the ratio $V_\mathrm{2kpc}/V_\mathrm{Rlast}$, which provides a measure of how rapidly the rotation curve rises, with values closer to unity corresponding to more rapidly rising rotation curves.

The goal of this work is not to establish statistically universal trends, but to investigate the range of structural and kinematic diversity that naturally emerges within our constrained Local Group simulations, and to explore the physical mechanisms associated with this diversity.

\subsubsection{Correlations with baryonic and dark matter structure} \label{sec:results:size}

We first examine the relation between rotation curve shape and galaxy size. The effective radius is a well-studied observable and is available for large samples of galaxies, including the SPARC dataset, enabling a direct comparison with observations. 

In the left panel of \Fig{fig:RCcoef-reff} we show $V_\mathrm{2kpc}/V_\mathrm{Rlast}$ as a function of effective radius for all \texttt{NIVARIA-LG} galaxies, coloured by stellar mass. For reference, SPARC galaxies are shown as grey diamonds, displaying good overall agreement with the simulated population. A subset of \texttt{NIVARIA-LG} systems populates the region of large effective radii and slowly rising rotation curves; this regime has proven challenging to reproduce in previous simulation efforts \citep{Cruz2025}, but emerges naturally in \texttt{NIVARIA-LG}. 
\Fig{fig:RCcoef-reff} shows that more extended galaxies tend to have more slowly rising rotation curves, with a Spearman coefficient of $r_\mathrm{s}=-0.56$.


Continuing with the baryonic component, and motivated by the illustrative examples discussed above, we now examine the spatial distribution of the gas. The central panel of \Fig{fig:RCcoef-reff} shows the rotation curve shape parameter as a function of the gas extent, $r_\mathrm{gas,90}$, defined as the radius enclosing 90$\%$ of the total gas mass of the galaxy. In this case, only \texttt{NIVARIA-LG} galaxies are shown, coloured by their stellar mass. 
A clear trend is present: galaxies with more extended gas distributions tend to exhibit more slowly rising rotation curves, and vice versa. The relation closely resembles that found for the effective radius, but is noticeably tighter, with $r_\mathrm{s}=-0.75$. 




Having explored the connection between rotation curve diversity and the baryonic distribution of galaxies, we now turn to the dark matter component. In the right panel of \Fig{fig:RCcoef-reff} we show the rotation curve shape parameter as a function of the inner slope of the dark matter density profile, $\alpha_{23}$, measured between 2 and 3 per cent of the halo virial radius, $R_{200}$ (see \Sec{sec:methods:DMprofiles}). More negative values of $\alpha_{23}$ correspond to steeper, cusp-like density profiles, while values closer to zero indicate shallower central density distributions. 

Again, there is a clear trend visible, with galaxies hosted by halos with shallower inner dark matter profiles tending to exhibit more slowly rising rotation curves. The Spearman correlation coefficient of $r_\mathrm{s}=-0.70$ indicates a moderate-to-strong correlation between the two quantities. To account for resolution effects, galaxies for which 2 per cent of $R_\mathrm{200}$ lies below the softening length of the simulation $\epsilon_\mathrm{DM}$ are shown in light gray in this panel. These represent a small fraction of the sample, and excluding them does not affect the observed trends.

\begin{figure}
\centering
  \includegraphics[width=8.8cm]{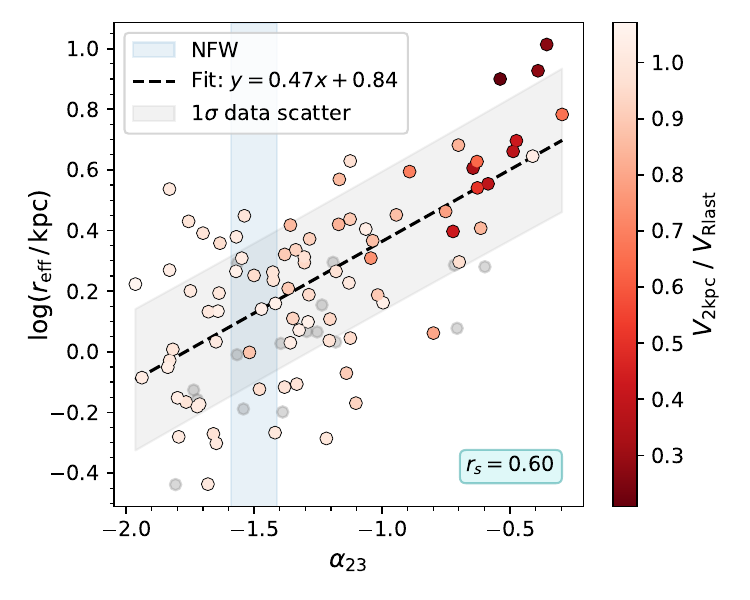}
  \vspace*{-0.4cm}
  \caption{Effective radius, $r_\mathrm{eff}$, as a function of the inner dark matter density profile slope, $\alpha_{23}$. Points are colored by the rotation curve shape parameter, $V_\mathrm{2kpc}/V_\mathrm{Rlast}$. The dashed line and gray shaded region denote the best-fit linear relation and its corresponding $1\sigma$ scatter. The vertical blue shaded region indicates the expected range for a pristine NFW halo ($\alpha_{23} \in [-1.59,1.41]$) assuming typical halo concentrations for halo mass in the range $10^9-10^{11} M_\odot$. Fully transparent gray points indicate galaxies excluded by our resolution limits.}
  \label{fig:RC-alpha-reff}
\end{figure}

\subsubsection{Disentangling the correlations} \label{sec:results:relation}

The correlations presented above raise an important question: are the relations between rotation curve shape, gas extent, and inner dark matter slope independent, or do they reflect a common underlying connection? 
To disentangle these relations, we perform a partial correlation analysis, controlling for stellar mass and for the mutual correlations between the structural parameters. Using the Pingouin python package \citep{Vallat2018}, we compute partial Spearman correlation coefficients, which quantify the correlation between two variables while controlling for one or more additional variables. We calculate this coefficient between the rotation curve shape parameter ($V_\mathrm{2kpc}/V_\mathrm{Rlast}$) and each individual structural property ($r_\mathrm{eff}$, $r_\mathrm{gas,90}$, and $\alpha_{23}$). We perform this analysis 
 first controlling only for the stellar mass ($M_\star$), and 
subsequently controlling for stellar mass alongside the other two remaining structural parameters. The results are shown in \Tab{tab:partial_correlations}.


\begin{table}
\centering
\caption{Partial Spearman correlation coefficients, $r$, and associated $p$-values between the rotation curve shape parameter ($V_\mathrm{2kpc}/V_\mathrm{Rlast}$) and individual galaxy structural properties. The first block shows correlations controlling only for stellar mass, while the second block additionally controls for the two remaining structural parameters.}
\label{tab:partial_correlations}

\begin{tabular}{lcc cc}
\hline
 & \multicolumn{2}{c}{Control: $M_\star$} & \multicolumn{2}{c}{Full control} \\
\cline{2-3} \cline{4-5}
$x_i$ & $r$ & $p$-value & $r$ & $p$-value \\
\hline
$r_\mathrm{eff}$    & -0.37 & $1.6 \times 10^{-4}$  & -0.02 & 0.81 \\
$r_\mathrm{gas,90}$ & -0.69 & $2.6 \times 10^{-15}$ & -0.64 & $1.43 \times 10^{-12}$ \\
$\alpha_{23}$       & -0.54 & $5.9 \times 10^{-9}$  & -0.46 & $2.59 \times 10^{-6}$ \\
\hline
\end{tabular}
\end{table}

\begin{figure*}
\centering
  \includegraphics[width=17.1cm]{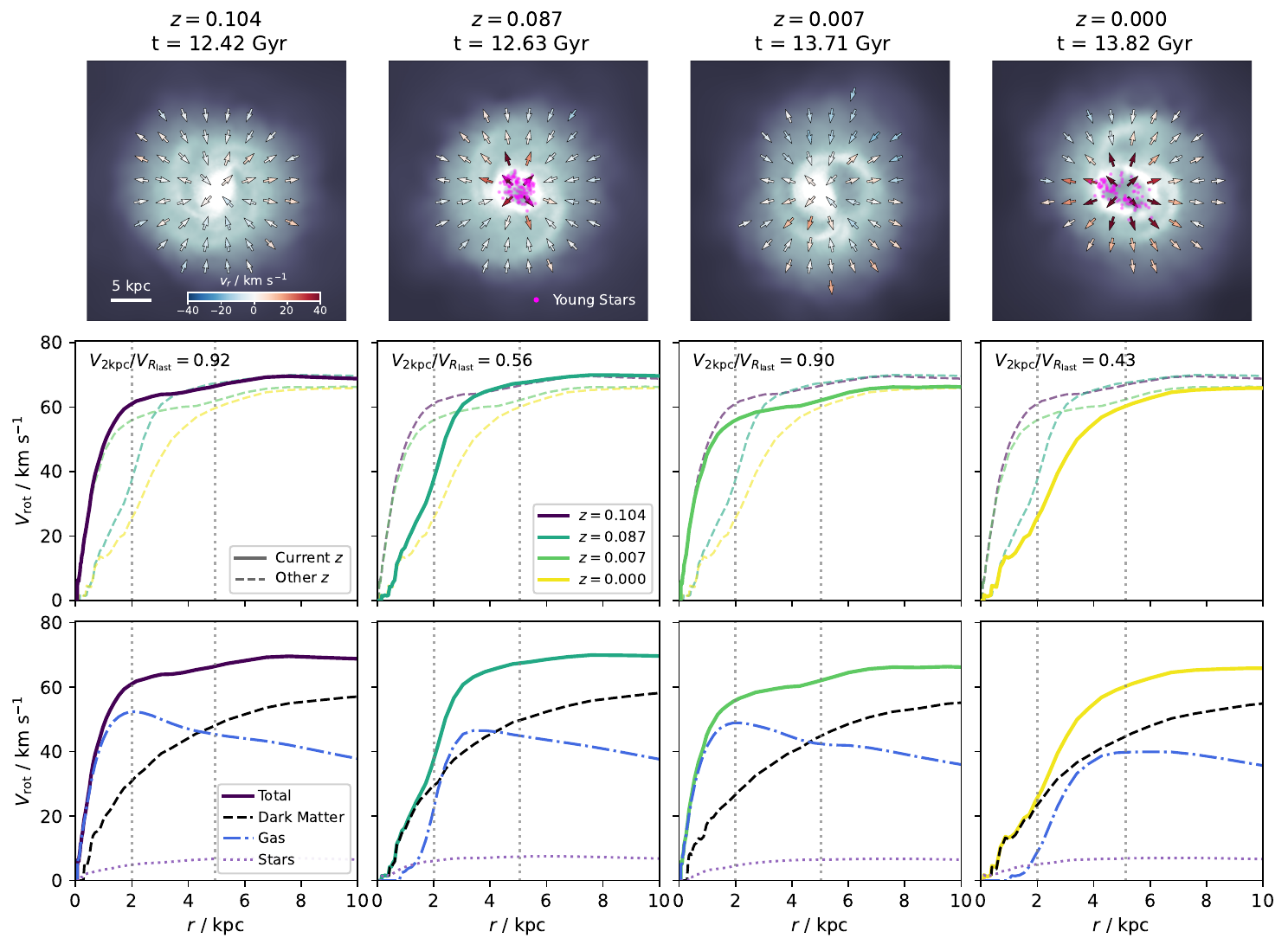}
  \caption{Evolution of Halo 35 across four different snapshots, with redshift and time indicated on top. Top row, face-on gas density map, with gas radial velocity indicated as coloured arrows (direction of the arrows indicates inflow or outflow, colour indicates the magnitude). Stars younger than 50 Myr are depicted as magenta dots. Middle row, full rotation curve, solid line indicates the given snapshot, the rest of the snapshots are shown as dashed lines, to ease the comparison. The rotation curve shape parameter, $V_\mathrm{2kpc}/V_\mathrm{Rlast}$, is indicated for each curve in the upper left corner (vertical dotted lines indicate $r=2$ kpc and $R_\mathrm{last}$). Bottom row, contribution from each of the different components to the full rotation curve at each snapshot, dark matter in dashed black lines, gas in dash-dotted blue and stars in dotted purple.}
  \label{fig:Halo35_RC_evol}
\end{figure*}
The statistical trends reveal a notable hierarchy among these internal properties. When controlling only for stellar mass, all three parameters display statistically significant correlations with the rotation curve shape, with the gas extent showing the tightest relation. However, when controlling to account for mutual dependencies, the independent correlation between the stellar effective radius ($r_\mathrm{eff}$) and the rotation curve shape largely disappears ($r = -0.02$, $p = 0.81$). Conversely, both the gas extent ($r = -0.64$) and the inner dark matter slope ($r = -0.46$) maintain strong and moderate independent correlations with the inner kinematics. These results suggest that within the framework of our simulations, the spatial extent of the stellar component may not act as an independent physical driver of the rotation curve shape. Instead, the moderate correlation between $r_\mathrm{eff}$ and $V_\mathrm{2kpc}/V_\mathrm{Rlast}$ highlighted in \Sec{sec:results:size} appears to be largely indirect, tracing variations that occur more fundamentally within the gas and dark matter components. 
To further explore the relation between these features, \Fig{fig:RC-alpha-reff} shows the relation between the $r_\mathrm{eff}$ and $\alpha_{23}$ for all \texttt{NIVARIA-LG} galaxies.  

The two quantities exhibit a clear overall trend, such that more extended stellar systems tend to reside in halos with shallower inner dark matter density profiles. For reference, we also indicate with a blue shaded region the expected range for $\alpha_{23}$ for a pristine NFW halo in the halo mass range $10^9-10^{11} M_\odot$. The concentration is derived from \citet[][eqs. 7 and 8]{DuttonMaccio2014} concentration-mass relation, and the value of the logarithmic slope is computed at $0.025R_{200}$.  We present the best linear fit as a dashed black line, together with the corresponding $1\sigma$ scatter region. The distribution of galaxies around this relation is not uniform, with increased scatter towards steeper inner density profiles. 
The rotation curve shape also varies systematically across this relation (see colourbar), reinforcing the picture that stellar extent is not an independent driver of the inner kinematics but is connected to broader variations in the gas and dark matter components.
The correlations above identify the gas distribution and inner dark matter structure as the properties most closely associated with rotation curve shape. Since both components can evolve over time through gas accretion, star formation, and feedback, we next ask how rotation curve shape itself evolves within individual galaxies.

\subsubsection{Rotation curve evolution}  \label{sec:results:origin:MAH}

The structural properties discussed above describe the present-day state of the galaxies. However, the gas distribution and star formation activity can change substantially over time, and these changes can modify the inner mass distribution and rotation curve. Thus, galaxies with similar global properties may exhibit different rotation curves simply because they are observed at different stages of their evolution.

To illustrate this, \Fig{fig:Halo35_RC_evol} follows the evolution of Halo 35 through four representative stages. The upper panels show face-on gas density maps, with the gas radial velocity field shown as coloured arrows, whose direction indicates the direction of the gas (inflow or outflow), and stars younger than 50 Myr overlaid as magenta dots. The middle panels show the corresponding rotation curves as solid lines, compared to the other three snapshots depicted as dashed lines. Before the first recent episode of enhanced star formation (first column), the gas is relatively concentrated towards the centre and the rotation curve rises rapidly. During the subsequent star formation episode (second column), the central gas distribution becomes depleted and the gas develops predominantly outward radial motions, coinciding with a more slowly rising rotation curve. Following this episode, the central gas distribution and rotation curve recover towards a more rapidly rising configuration (third column). A second episode of enhanced star formation occurs later (last column, $z=0$), when the galaxy again shows a depleted central gas distribution and a slowly rising rotation curve. 

The bottom row in \Fig{fig:Halo35_RC_evol} shows the contribution from the different components to the full rotation curve at each snapshot. We see that, while the dark matter contribution (dashed black lines) remains similar across time, the gas component (dash-dotted blue lines) is changing significantly in the central region while reaching similar values at $r>R_\mathrm{last}$, thus affecting the full rotation curve. This shows that not only the total amount but the distribution of the gas plays a key role in shaping the rotation curve of a galaxy, which can explain the presence of slowly rising rotation curves with significant baryonic contributions discussed by \citet{Santos-Santos2020}.

\begin{figure}
\centering
  \includegraphics[width=7.4cm]{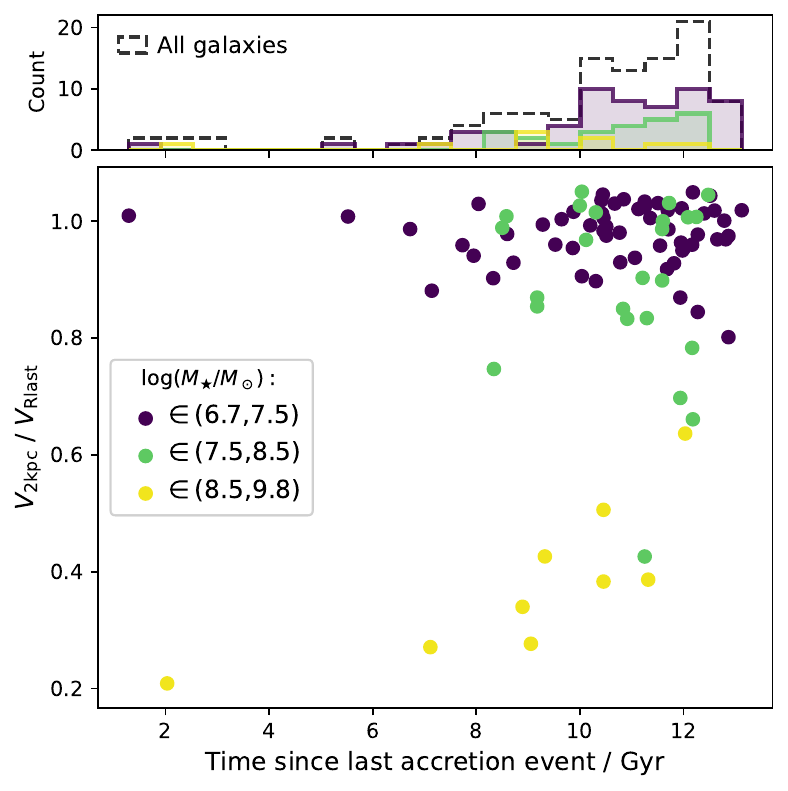}
  \caption{Rotation curve shape parameter $V_\mathrm{2kpc}/V_\mathrm{Rlast}$ as a function of the time since the last significant accretion event, defined as a total mass increase of at least 20 per cent in 150 Myr. Colours indicate increasing mass bins, as indicated in the legend. The panel on top shows the distribution of the recent accretion timescales for all galaxies (dashed black lines) and for each bin (solid coloured lines).}
  \label{fig:RC-mergers}
\end{figure}

This example shows how the rotation curve shape evolves within an individual galaxy. Changes in the gas distribution associated with star formation and feedback modify the central gravitational potential, producing different rotation curve shapes at different times. The diversity seen across galaxies at $z=0$ may therefore partly reflect differences in their evolutionary state.

An important question is what regulates these cycles. Recent accretion events and mergers can redistribute gas and stars, alter their angular momentum, and trigger episodes of enhanced star formation \citep[e.g.][]{Starkenburg2016,Rey2020,Wright2025}. We therefore investigate whether the rotation curve shape is related to the recent assembly history of the \texttt{NIVARIA-LG} galaxies. Using the AHF merger trees, we define a significant accretion event as an increase in total mass (dark matter, stars, and gas) of at least 20 per cent within 150 Myr. \Fig{fig:RC-mergers} shows the rotation curve shape parameter, $V_\mathrm{2kpc}/V_\mathrm{Rlast}$, as a function of the time since the most recent accretion event for \texttt{NIVARIA-LG} galaxies separated into stellar-mass bins. We exclude 10 galaxies with incomplete or noisy merger trees, leaving a final sample of 94 systems.

The upper panel shows the distribution of times since the last accretion event for the full sample and for each stellar mass bin. For the two lower-mass bins ($M_\star < 10^{8.5},M_\odot$), we find no clear relation between rotation curve shape and recent assembly history. In contrast, the most massive galaxies show a trend in which more recent accretion events are associated with more slowly rising rotation curves. This suggests that recent accretion may contribute to modifying the internal structure of the more massive galaxies, potentially by redistributing gas and altering the central mass distribution. The absence of a clear trend at lower masses may reflect the stronger role of stochastic internal feedback processes. The substantial scatter indicates, however, that recent accretion is not sufficient to determine the rotation curve shape. The response to an accretion event depends on its mass, orbit, gas content, and subsequent feedback, while gas can also reach the central regions through continuous accretion and recycling. The constrained Local Group environment may further influence these processes by regulating the supply and redistribution of gas. In particular, \citet{Arora2022} found that Local Group dwarf galaxies have different gas distributions from isolated dwarfs, suggesting that environmental conditions may contribute to shaping the baryon cycle and, indirectly, the diversity of rotation curves.

Overall, the results suggest that rotation curve diversity reflects both differences in present-day structure and differences in evolutionary state. The gas distribution is closely linked to the rotation curve shape, while its evolution through accretion, star formation, and feedback can produce changes in the inner potential over time. 

\section{Conclusions} \label{sec:conclusions}


In this pilot paper we have presented the first results from \texttt{NIVARIA-LG}, an intermediate-resolution simulation serving as the first feasibility study of the broader NIVARIA project, whose full high-resolution simulation campaign is yet to be carried out (PI A. Di Cintio).
\texttt{NIVARIA-LG} is a constrained cosmological hydrodynamical simulation of the Local Group and its surrounding environment, designed to study the evolution of dwarf galaxies in a realistic cosmological setting. Analysing a sample of non-satellite dwarf galaxies with stellar masses $6\times10^6 \lesssim M_\star/M_\odot \lesssim 10^{10}$, we investigated the physical origin of the diversity of dwarf galaxy rotation curve shapes. By comparing  simulations directly with   observational galaxy datasets, we explored the structural and evolutionary processes that regulate the inner kinematics of dwarf galaxies. 
Our main findings are summarised as follows.\\\\

The \texttt{NIVARIA-LG} simulation:
\begin{itemize}
    \item  produces a realistic dwarf galaxy population, reproducing abundance matching relations, the observed luminosity function of the Local Group, and the expected Baryonic Tully-Fisher relation (\Fig{fig:mstar-mhalo});
    
    \item  provides a good match to the observed stellar mass--size relation over a broad range of effective radii. Compared to previous simulations, \texttt{NIVARIA-LG} better reproduces the observed diversity of galaxy sizes, reaching both very extended galaxies (large $r_{\rm eff}$) and compact dwarfs (small $r_{\rm eff}$) (\Fig{fig:size-mstar} and \Tab{tab:mass_radius_bins});
    
    \item naturally generates a continuous distribution of rotation curve shapes, from steeply to slowly rising profiles, in better agreement with the observed SPARC diversity of rotation curve shapes than any previous simulation suites (\Fig{fig:RCs-comparison}), thanks to its realistic and rich  Group environment that  influences and  regulates gas accretion and to the matter redistribution given by the feedback prescriptions used;
    
    \item indicates that the spatial extent of the gas component, quantified by $r_{\rm gas,90}$, shows the strongest association with rotation curve shape: galaxies with more extended gas distributions systematically exhibit more slowly rising rotation curves. Similarly, the inner dark matter density slope, $\alpha_{23}$, also plays a key role: galaxies with shallower central density profiles have systematically lower inner circular velocities, supporting the importance of feedback-driven core formation as additional contribution to rotation curves diversity (\Fig{fig:example-bin2}, \Fig{fig:RCcoef-reff} and \Tab{tab:partial_correlations});
    
    \item the stellar effective radius, $r_\mathrm{eff}$, correlates with rotation curve shape when controlling only for stellar mass, but this correlation largely disappears when also controlling for the gas and dark matter-related parameters (\Tab{tab:partial_correlations}). This suggests that $r_\mathrm{eff}$ alone is not an independent predictor of rotation curve diversity, but instead traces broader structural transformations involving the baryonic and dark matter components. Indeed, we find a clear linear trend between $\log(r_\mathrm{eff})$ and the inner dark matter slope $\alpha_{23}$ of dwarf galaxies, as expected in models in which both dark matter and stars are redistributed  during bursty star formation driving gas outflows (\Fig{fig:RC-alpha-reff});
    

    \item highlights that the shape of the rotation curve of a galaxy reflects the stage of the galaxy in its baryonic feedback cycle, regulated by gas accretion, star formation and SN feedback. Galaxies with similar global properties can be in different stages of their evolution and hence show very distinct rotation curves (\Fig{fig:Halo35_RC_evol});

    \item shows that late-time accretion can contribute to further modify the central mass distribution, specially at the high-mass end, i.e. galaxies with $M_\star>10^{8.5},M_\odot$, in which recent significant accretion events are associated with more slowly rising rotation curves, although with substantial scatter (\Fig{fig:RC-mergers}).

\end{itemize}

The ability of the \texttt{NIVARIA-LG} simulations to naturally reproduce the full range of observed rotation curve shapes addresses a challenge faced by many state-of-the-art cosmological simulations. While \citet{Jackson2025} quantify baryonic feedback by measuring the central dark matter mass reduction relative to dark-matter-only counterparts, our results offer a complementary perspective by linking kinematic diversity directly to internal, potentially observable structural properties. 
Although previous works have argued that part of the observed diversity may arise from observational systematics \citep[e.g.][]{Santos-Santos2020,Roper2023}, our analysis assumes that the diversity is physical and instead focuses on identifying its underlying drivers. 
Our results  differ from those of  \citet{Cruz2025}, who argue that overly strong feedback in previous simulations may prevent the formation of compact dwarf galaxies. In \texttt{NIVARIA-LG}, the same feedback prescription used in \texttt{NIHAO} produces both compact and extended dwarfs, as well as a broad diversity of rotation curves. This suggests that the feedback prescription alone does not determine the diversity of dwarf galaxy sizes and inner kinematics, and that the cosmological and environmental context may also play an important role. 

In this context, the strong connection we find between gas extent and inner kinematics is particularly interesting in light of the results of \citet{Arora2022}, who showed that Local Group dwarf galaxies possess systematically different gas distributions from the general dwarf population. This may be particularly relevant for \texttt{NIVARIA-LG}, which follows galaxies in a constrained Local Group-like environment rather than in isolation. The environmental conditions can affect the supply, redistribution, and recycling of gas, potentially changing the frequency and strength of star formation and feedback episodes. In this picture, the environment does not directly set the rotation curve shape, but can regulate the baryon cycle that ultimately determines the central gas and dark matter distribution. This provides a possible explanation for why \texttt{NIVARIA-LG} is able to reproduce a broad diversity of rotation curves while simulations of isolated galaxies may produce a more restricted range.

This work represents the first step and feasibility study of the broader NIVARIA project, showcasing the power of combining constrained Local Group initial conditions with state-of-the-art galaxy formation prescriptions. Future studies will make use of higher-resolution simulations of the suite to improve our understanding of the connection between baryons and dark matter in dwarf galaxies. We will also explore the most extreme systems produced in the current simulations, in particular those with the largest extents and slowly rising rotation curves, to better understand their origin. In addition, we will analyse simulations with the supermassive black hole model in \texttt{Gasoline2} enabled, using otherwise identical initial conditions, in order to isolate the impact of black holes and AGN feedback on dwarf galaxy evolution.


\begin{acknowledgements}
    The authors thanks J. Navarro, I. Santos-Santos, and E. Arjona-G\'alvez for useful discussions, J. Benavides for providing DESI data, and the FIRE collaboration for making their data publicly available. ACS, ADC and GGB acknowledge financial support from the Spanish Ministry of Science and Innovation (MICINN), Consolidación Investigadora program, CNS2023-144669, project ``Tiny'' (PI A. Di Cintio), and the 2024 call ``Proyectos de Generación de Conocimiento'', grant number PID2024-160009NA-I00, proyecto ``INGENIO''. SCB acknowledges financial support from the (MICINN) through RYC2022-035838-I, PID2021-128131NB- I00 (CoBEARD) and CNS2022-135482 projects. AK is supported by the Agencia Estatal de Investigación (AEI, Spain)  under project  PID2024-156100NB-C21 financed by MICIU /AEI/10.13039/501100011033 / FEDER, UE. He further thanks Iron Maiden for `fear of the dark'.

\end{acknowledgements}


\bibliographystyle{aa}
\bibliography{archive}

@ARTICLE{Muni2025,
       author = {{Muni}, Claudia and {Pontzen}, Andrew and {Read}, Justin I. and {Agertz}, Oscar and {Rey}, Martin P. and {Taylor}, Ethan and {Kim}, Stacy Y. and {Gray}, Emily I.},
        title = "{EDGE: dark matter core creation depends on the timing of star formation}",
      journal = {\mnras},
         year = 2025,
        month = jan,
       volume = {536},
       number = {1},
        pages = {314-323},
          doi = {10.1093/mnras/stae2748},
archivePrefix = {arXiv},
       eprint = {2407.14579},
 primaryClass = {astro-ph.GA},
       adsurl = {https://ui.adsabs.harvard.edu/abs/2025MNRAS.536..314M}
}

@ARTICLE{DiCintio2017,
       author = {{Di Cintio}, Arianna and {Brook}, Chris B. and {Dutton}, Aaron A. and {Macci{\`o}}, Andrea V. and {Obreja}, Aura and {Dekel}, Avishai},
        title = "{NIHAO - XI. Formation of ultra-diffuse galaxies by outflows}",
      journal = {\mnras},
         year = 2017,
        month = mar,
       volume = {466},
       number = {1},
        pages = {L1-L6},
          doi = {10.1093/mnrasl/slw210},
archivePrefix = {arXiv},
       eprint = {1608.01327},
 primaryClass = {astro-ph.GA},
       adsurl = {https://ui.adsabs.harvard.edu/abs/2017MNRAS.466L...1D}
}

@ARTICLE{Sarrato-Alos2026,
       author = {{Sarrato-Al{\'o}s}, J. and {Bullock}, J. and {Di Cintio}, A. and {Brook}, C. and {Valenciano}, F. and {Macci{\`o}}, A.~V.},
        title = "{The impact of star formation histories on the inner dark matter density slopes of galaxies}",
      journal = {\aap},
         year = 2026,
        month = jun,
       volume = {710},
          eid = {A95},
        pages = {A95},
          doi = {10.1051/0004-6361/202558718},
archivePrefix = {arXiv},
       eprint = {2601.17113},
 primaryClass = {astro-ph.GA},
       adsurl = {https://ui.adsabs.harvard.edu/abs/2026A&A...710A..95S}
}

@ARTICLE{PontzenGovernato2014,
       author = {{Pontzen}, Andrew and {Governato}, Fabio},
        title = "{Cold dark matter heats up}",
      journal = {\nat},
         year = 2014,
        month = feb,
       volume = {506},
       number = {7487},
        pages = {171-178},
          doi = {10.1038/nature12953},
archivePrefix = {arXiv},
       eprint = {1402.1764},
 primaryClass = {astro-ph.CO},
       adsurl = {https://ui.adsabs.harvard.edu/abs/2014Natur.506..171P}
}

@ARTICLE{Arora2022,
       author = {{Arora}, Nikhil and {Macci{\`o}}, Andrea V. and {Courteau}, St{\'e}phane and {Buck}, Tobias and {Libeskind}, Noam I. and {Sorce}, Jenny G. and {Brook}, Chris B. and {Hoffman}, Yehuda and {Yepes}, Gustavo and {Carlesi}, Edoardo and {Stone}, Connor},
        title = "{NIHAO-LG: the uniqueness of Local Group dwarf galaxies}",
      journal = {\mnras},
         year = 2022,
        month = jun,
       volume = {512},
       number = {4},
        pages = {6134-6149},
          doi = {10.1093/mnras/stac893},
archivePrefix = {arXiv},
       eprint = {2109.07487},
 primaryClass = {astro-ph.GA},
       adsurl = {https://ui.adsabs.harvard.edu/abs/2022MNRAS.512.6134A}
}

@ARTICLE{Wright2025,
       author = {{Wright}, Anna C. and {Brooks}, Alyson M. and {Tremmel}, Michael and {Young}, Jason E. and {Munshi}, Ferah and {Quinn}, Tom},
        title = "{The Merger-Driven Formation of Classical Low Surface Brightness Galaxies in Romulus25}",
      journal = {arXiv e-prints},
         year = 2025,
        month = jul,
          eid = {arXiv:2507.21231},
        pages = {arXiv:2507.21231},
          doi = {10.48550/arXiv.2507.21231},
archivePrefix = {arXiv},
       eprint = {2507.21231},
 primaryClass = {astro-ph.GA},
       adsurl = {https://ui.adsabs.harvard.edu/abs/2025arXiv250721231W}
}

@ARTICLE{Gannon2024,
       author = {{Gannon}, Jonah S. and {Ferr{\'e}-Mateu}, Anna and {Forbes}, Duncan A. and {Brodie}, Jean P. and {Buzzo}, Maria Luisa and {Romanowsky}, Aaron J.},
        title = "{A Catalogue and analysis of ultra-diffuse galaxy spectroscopic properties}",
      journal = {\mnras},
         year = 2024,
        month = jun,
       volume = {531},
       number = {1},
        pages = {1856-1869},
          doi = {10.1093/mnras/stae1287},
archivePrefix = {arXiv},
       eprint = {2405.09104},
 primaryClass = {astro-ph.GA},
       adsurl = {https://ui.adsabs.harvard.edu/abs/2024MNRAS.531.1856G}
}

@ARTICLE{Benavides2026,
       author = {{Benavides}, Jose and {Sales}, Laura V. and {Navarro}, Julio F. and {White}, Simon D.~M. and {Frenk}, Carlos S. and {Oman}, Kyle A. and {Cole}, Shaun},
        title = "{The Abundance of Thin Dwarf Galaxies: a Challenge for Cosmological Simulations}",
      journal = {The Open Journal of Astrophysics},
         year = 2026,
        month = may,
       volume = {9},
        pages = {62091},
          doi = {10.33232/001c.162091},
archivePrefix = {arXiv},
       eprint = {2512.11035},
 primaryClass = {astro-ph.GA},
       adsurl = {https://ui.adsabs.harvard.edu/abs/2026OJAp....962091B}
}

@ARTICLE{DESI2026,
       author = {{DESI Collaboration} and {Abdul Karim}, M. and {Adame}, A.~G. and {Aguado}, D. and {Aguilar}, J. and {Ahlen}, S. and {Alam}, S. and {Aldering}, G. and {Alexander}, D.~M. and {Alfarsy}, R. and {Allen}, L. and {Allende Prieto}, C. and {Alves}, O. and {Anand}, A. and {Andrade}, U. and {Armengaud}, E. and {Avila}, S. and {Aviles}, A. and {Awan}, H. and {Bailey}, S. and {Baleato Lizancos}, A. and {Ballester}, O. and {Bault}, A. and {Bautista}, J. and {Bean}, R. and {Behera}, J. and {BenZvi}, S. and {Beraldo e Silva}, L. and {Bermejo-Climent}, J.~R. and {Beutler}, F. and {Bianchi}, D. and {Blake}, C. and {Blum}, R. and {Bolton}, A.~S. and {Bonici}, M. and {Brieden}, S. and {Brodzeller}, A. and {Brooks}, D. and {Buckley-Geer}, E. and {Burtin}, E. and {Bystr{\"o}m}, A. and {Canning}, R. and {Carnero Rosell}, A. and {Carr}, A. and {Carrilho}, P. and {Casas}, L. and {Castander}, F.~J. and {Cereskaite}, R. and {Cervantes-Cota}, J.~L. and {Chaussidon}, E. and {Chaves-Montero}, J. and {Chen}, S. and {Chen}, X. and {Circosta}, C. and {Claybaugh}, T. and {Cole}, S. and {Cooper}, A.~P. and {Cousinou}, M.-C. and {Cuceu}, A. and {Davis}, T.~M. and {Dawson}, K.~S. and {de Belsunce}, R. and {de la Cruz}, R. and {de la Macorra}, A. and {de Mattia}, A. and {Deiosso}, N. and {Della Costa}, J. and {Demina}, R. and {Demirbozan}, U. and {DeRose}, J. and {Dey}, A. and {Dey}, B. and {Ding}, J. and {Ding}, Z. and {Doel}, P. and {Douglass}, K. and {Dowicz}, M. and {Ebina}, H. and {Edelstein}, J. and {Eisenstein}, D.~J. and {Elbers}, W. and {Emas}, N. and {Escoffier}, S. and {Fagrelius}, P. and {Fan}, X. and {Fanning}, K. and {Favole}, G. and {Fawcett}, V.~A. and {Fern{\'a}ndez-Garc{\'\i}a}, E. and {Ferraro}, S. and {Findlay}, N. and {Font-Ribera}, A. and {Forero-Romero}, J.~E. and {Forero-S{\'a}nchez}, D. and {Frenk}, C.~S. and {G{\"a}nsicke}, B.~T. and {Galbany}, L. and {Garc{\'\i}a-Bellido}, J. and {Garcia-Quintero}, C. and {Garrison}, L.~H. and {Gazta{\~n}aga}, E. and {Gil-Mar{\'\i}n}, H. and {Gloudemans}, A. and {Gnedin}, O.~Y. and {Gontcho A Gontcho}, S. and {Gonzalez}, D. and {Gonzalez-Morales}, A.~X. and {Gonzalez-Perez}, V. and {Gordon}, C. and {Graur}, O. and {Green}, D. and {Gruen}, D. and {Gsponer}, R. and {Guandalin}, C. and {Gutierrez}, G. and {Guy}, J. and {Hahn}, C. and {Han}, J.~J. and {Han}, J. and {He}, S. and {Herrera-Alcantar}, H.~K. and {Heydenreich}, S. and {Honscheid}, K. and {Hou}, J. and {Howlett}, C. and {Huterer}, D. and {Ir{\v{s}}i{\v{c}}}, V. and {Ishak}, M. and {Jacques}, A. and {Jiang}, L. and {Jimenez}, J. and {Jing}, Y.~P. and {Joachimi}, B. and {Joudaki}, S. and {Joyce}, R. and {Jullo}, E. and {Juneau}, S. and {Kara{\c{c}}ayl{\i}}, N.~G. and {Karim}, T. and {Kehoe}, R. and {Kent}, S. and {Khederlarian}, A. and {Kirkby}, D. and {Kisner}, T. and {Kitaura}, F.-S. and {Kizhuprakkat}, N. and {Kong}, H. and {Koposov}, S.~E. and {Kremin}, A. and {Krolewski}, A. and {Lahav}, O. and {Lai}, Y. and {Lamman}, C. and {Lan}, T.-W. and {Landriau}, M. and {Lang}, D. and {Lange}, J.~U. and {Lasker}, J. and {Le Goff}, J.~M. and {Le Guillou}, L. and {Leauthaud}, A. and {Levi}, M.~E. and {Li}, S. and {Li}, T.~S. and {Liu}, W. and {Lodha}, K. and {Lokken}, M. and {Luo}, Y. and {Magneville}, C. and {Manera}, M. and {Manser}, C.~J. and {Margala}, D. and {Martini}, P. and {Maus}, M. and {McCullough}, J. and {McDonald}, P. and {Medina}, G.~E. and {Medina-Varela}, L. and {Meisner}, A. and {Mena-Fern{\'a}ndez}, J. and {Menegas}, A. and {Meneses-Rizo}, J. and {Mezcua}, M. and {Miquel}, R. and {Montero-Camacho}, P. and {Moon}, J. and {Moustakas}, J. and {Mu{\~n}oz-Guti{\'e}rrez}, A. and {Mu noz-Santos}, D. and {Myers}, A.~D. and {Myles}, J. and {Nadathur}, S. and {Najita}, J. and {Napolitano}, L. and {Newman}, J.~A. and {Nikakhtar}, F. and {Nikutta}, R. and {Niz}, G. and {Noriega}, H.~E. and {Nugent}, P.},
        title = "{Data Release 1 of the Dark Energy Spectroscopic Instrument}",
      journal = {\aj},
         year = 2026,
        month = may,
       volume = {171},
       number = {5},
          eid = {285},
        pages = {285},
          doi = {10.3847/1538-3881/ae4c43},
archivePrefix = {arXiv},
       eprint = {2503.14745},
 primaryClass = {astro-ph.CO},
       adsurl = {https://ui.adsabs.harvard.edu/abs/2026AJ....171..285D}
}

@ARTICLE{Starkenburg2016,
       author = {{Starkenburg}, T.~K. and {Helmi}, A. and {Sales}, L.~V.},
        title = "{Dark influences II. Gas and star formation in minor mergers of dwarf galaxies with dark satellites}",
      journal = {\aap},
         year = 2016,
        month = mar,
       volume = {587},
          eid = {A24},
        pages = {A24},
          doi = {10.1051/0004-6361/201527247},
archivePrefix = {arXiv},
       eprint = {1508.06281},
 primaryClass = {astro-ph.GA},
       adsurl = {https://ui.adsabs.harvard.edu/abs/2016A&A...587A..24S}
}

@ARTICLE{Rey2020,
       author = {{Rey}, Martin P. and {Pontzen}, Andrew and {Agertz}, Oscar and {Orkney}, Matthew D.~A. and {Read}, Justin I. and {Rosdahl}, Joakim},
        title = "{EDGE: from quiescent to gas-rich to star-forming low-mass dwarf galaxies}",
      journal = {\mnras},
         year = 2020,
        month = sep,
       volume = {497},
       number = {2},
        pages = {1508-1520},
          doi = {10.1093/mnras/staa1640},
archivePrefix = {arXiv},
       eprint = {2004.09530},
 primaryClass = {astro-ph.GA},
       adsurl = {https://ui.adsabs.harvard.edu/abs/2020MNRAS.497.1508R}
}

@ARTICLE{Savino2022,
       author = {{Savino}, Alessandro and {Weisz}, Daniel R. and {Skillman}, Evan D. and {Dolphin}, Andrew and {Kallivayalil}, Nitya and {Wetzel}, Andrew and {Anderson}, Jay and {Besla}, Gurtina and {Boylan-Kolchin}, Michael and {Bullock}, James S. and {Cole}, Andrew A. and {Collins}, Michelle L.~M. and {Cooper}, M.~C. and {Deason}, Alis J. and {Dotter}, Aaron L. and {Fardal}, Mark and {Ferguson}, Annette M.~N. and {Fritz}, Tobias K. and {Geha}, Marla C. and {Gilbert}, Karoline M. and {Guhathakurta}, Puragra and {Ibata}, Rodrigo and {Irwin}, Michael J. and {Jeon}, Myoungwon and {Kirby}, Evan and {Lewis}, Geraint F. and {Mackey}, Dougal and {Majewski}, Steven R. and {Martin}, Nicolas and {McConnachie}, Alan and {Patel}, Ekta and {Rich}, R. Michael and {Simon}, Joshua D. and {Sohn}, Sangmo Tony and {Tollerud}, Erik J. and {van der Marel}, Roeland P.},
        title = "{The Hubble Space Telescope Survey of M31 Satellite Galaxies. I. RR Lyrae-based Distances and Refined 3D Geometric Structure}",
      journal = {\apj},
         year = 2022,
        month = oct,
       volume = {938},
       number = {2},
          eid = {101},
        pages = {101},
          doi = {10.3847/1538-4357/ac91cb},
archivePrefix = {arXiv},
       eprint = {2206.02801},
 primaryClass = {astro-ph.GA},
       adsurl = {https://ui.adsabs.harvard.edu/abs/2022ApJ...938..101S}
}

@ARTICLE{McConnachie2012,
       author = {{McConnachie}, Alan W.},
        title = "{The Observed Properties of Dwarf Galaxies in and around the Local Group}",
      journal = {\aj},
         year = 2012,
        month = jul,
       volume = {144},
       number = {1},
          eid = {4},
        pages = {4},
          doi = {10.1088/0004-6256/144/1/4},
archivePrefix = {arXiv},
       eprint = {1204.1562},
 primaryClass = {astro-ph.CO},
       adsurl = {https://ui.adsabs.harvard.edu/abs/2012AJ....144....4M}
}

@ARTICLE{DuttonMaccio2014,
       author = {{Dutton}, Aaron A. and {Macci{\`o}}, Andrea V.},
        title = "{Cold dark matter haloes in the Planck era: evolution of structural parameters for Einasto and NFW profiles}",
      journal = {\mnras},
         year = 2014,
        month = jul,
       volume = {441},
       number = {4},
        pages = {3359-3374},
          doi = {10.1093/mnras/stu742},
archivePrefix = {arXiv},
       eprint = {1402.7073},
 primaryClass = {astro-ph.CO},
       adsurl = {https://ui.adsabs.harvard.edu/abs/2014MNRAS.441.3359D}
}

@ARTICLE{Governato2012,
       author = {{Governato}, F. and {Zolotov}, A. and {Pontzen}, A. and {Christensen}, C. and {Oh}, S.~H. and {Brooks}, A.~M. and {Quinn}, T. and {Shen}, S. and {Wadsley}, J.},
        title = "{Cuspy no more: how outflows affect the central dark matter and baryon distribution in {\ensuremath{\Lambda}} cold dark matter galaxies}",
      journal = {\mnras},
         year = 2012,
        month = may,
       volume = {422},
       number = {2},
        pages = {1231-1240},
          doi = {10.1111/j.1365-2966.2012.20696.x},
archivePrefix = {arXiv},
       eprint = {1202.0554},
 primaryClass = {astro-ph.CO},
       adsurl = {https://ui.adsabs.harvard.edu/abs/2012MNRAS.422.1231G}
}

@ARTICLE{Gannon2025,
       author = {{Gannon}, Jonah S. and {Di Cintio}, Arianna and {Forbes}, Duncan A. and {Garc{\'\i}a-Bethencourt}, Guacimara and {Brodie}, Jean P. and {Libeskind}, Noam and {Couch}, Warrick J. and {Hartke}, Johanna},
        title = "{Ultra-diffuse, ultra-different: observed versus simulated ultra-diffuse galaxies live in fundamentally different haloes}",
      journal = {\mnras},
         year = 2025,
        month = dec,
       volume = {544},
       number = {4},
        pages = {3094-3103},
          doi = {10.1093/mnras/staf1954},
archivePrefix = {arXiv},
       eprint = {2511.04006},
 primaryClass = {astro-ph.GA},
       adsurl = {https://ui.adsabs.harvard.edu/abs/2025MNRAS.544.3094G}
}

@ARTICLE{Garcia-Bethencourt2026,
       author = {{Garc{\'\i}a-Bethencourt}, Guacimara and {Di Cintio}, Arianna and {Comer{\'o}n}, S{\'e}bastien and {Arjona-G{\'a}lvez}, Elena and {Contreras-Santos}, Ana and {Cardona-Barrero}, Salvador and {Brook}, Chris B.~A. and {Negri}, Andrea and {Libeskind}, Noam I. and {Knebe}, Alexander},
        title = "{H I-bearing dark galaxy predictions from constrained Local Group simulations: How many and where to find them}",
      journal = {\aap},
         year = 2026,
        month = jun,
       volume = {710},
          eid = {A156},
        pages = {A156},
          doi = {10.1051/0004-6361/202558801},
archivePrefix = {arXiv},
       eprint = {2601.04024},
 primaryClass = {astro-ph.GA},
       adsurl = {https://ui.adsabs.harvard.edu/abs/2026A&A...710A.156G}
}

@ARTICLE{Dupuy2022,
       author = {{Dupuy}, Alexandra and {Libeskind}, Noam I. and {Hoffman}, Yehuda and {Courtois}, H{\'e}l{\`e}ne M. and {Gottl{\"o}ber}, Stefan and {Grand}, Robert J.~J. and {Knebe}, Alexander and {Sorce}, Jenny G. and {Tempel}, Elmo and {Tully}, R. Brent and {Vogelsberger}, Mark and {Wang}, Peng},
        title = "{Anisotropic satellite accretion on to the Local Group with HESTIA}",
      journal = {\mnras},
         year = 2022,
        month = nov,
       volume = {516},
       number = {3},
        pages = {4576-4584},
          doi = {10.1093/mnras/stac2486},
archivePrefix = {arXiv},
       eprint = {2208.14648},
 primaryClass = {astro-ph.GA},
       adsurl = {https://ui.adsabs.harvard.edu/abs/2022MNRAS.516.4576D}
}

@ARTICLE{Newton2023,
       author = {{Newton}, Oliver and {Di Cintio}, Arianna and {Cardona-Barrero}, Salvador and {Libeskind}, Noam I. and {Hoffman}, Yehuda and {Knebe}, Alexander and {Sorce}, Jenny G. and {Steinmetz}, Matthias and {Tempel}, Elmo},
        title = "{The Undiscovered Ultradiffuse Galaxies of the Local Group}",
      journal = {\apjl},
         year = 2023,
        month = apr,
       volume = {946},
       number = {2},
          eid = {L37},
        pages = {L37},
          doi = {10.3847/2041-8213/acc2bb},
archivePrefix = {arXiv},
       eprint = {2212.05066},
 primaryClass = {astro-ph.GA},
       adsurl = {https://ui.adsabs.harvard.edu/abs/2023ApJ...946L..37N}
}

@ARTICLE{Koudmani2025,
       author = {{Koudmani}, Sophie and {Rennehan}, Douglas and {Somerville}, Rachel S. and {Hayward}, Christopher C. and {Angl{\'e}s-Alc{\'a}zar}, Daniel and {Orr}, Matthew E. and {Sands}, Isabel S. and {Wellons}, Sarah},
        title = "{Diverse dark matter profiles in FIRE dwarfs: black holes, cosmic rays and the cusp─core enigma}",
      journal = {\mnras},
         year = 2025,
        month = jun,
       volume = {540},
       number = {2},
        pages = {1928-1950},
          doi = {10.1093/mnras/staf778},
archivePrefix = {arXiv},
       eprint = {2409.02172},
 primaryClass = {astro-ph.GA},
       adsurl = {https://ui.adsabs.harvard.edu/abs/2025MNRAS.540.1928K}
}

@ARTICLE{Jackson2025,
       author = {{Jackson}, R.~A. and {Navarro}, J.~F. and {Santos-Santos}, I.~M.~E. and {Kaviraj}, S. and {Yi}, S.~K. and {Peirani}, S. and {Dubois}, Y. and {Martin}, G. and {Devriendt}, J.~E.~G. and {Slyz}, A. and {Pichon}, C. and {Volonteri}, M. and {Kimm}, T. and {Kraljic}, K.},
        title = "{The diversity of rotation curves of galaxies in the NEWHORIZON cosmological simulation}",
      journal = {\mnras},
         year = 2025,
        month = jun,
       volume = {539},
       number = {4},
        pages = {3797-3807},
          doi = {10.1093/mnras/staf667},
       adsurl = {https://ui.adsabs.harvard.edu/abs/2025MNRAS.539.3797J}
}

@ARTICLE{Dado2026,
       author = {{Dado}, Diego and {Oman}, Kyle A. and {Harborne}, Katherine E. and {Fragkoudi}, Francesca and {Schaye}, Joop and {Schaller}, Matthieu and {Ben{\'\i}tez-Llambay}, Alejandro and {Chaikin}, Evgenii and {Frenk}, Carlos S. and {Hu{\v{s}}ko}, Filip and {Ploeckinger}, Sylvia and {Richings}, Alexander J.},
        title = "{Dynamical disequilibrium in dwarf galaxies: rethinking gas dynamics, rotation curves, and dark matter inference}",
      journal = {\mnras},
         year = 2026,
        month = apr,
       volume = {547},
       number = {2},
          eid = {stag356},
        pages = {stag356},
          doi = {10.1093/mnras/stag356},
archivePrefix = {arXiv},
       eprint = {2512.11033},
 primaryClass = {astro-ph.GA},
       adsurl = {https://ui.adsabs.harvard.edu/abs/2026MNRAS.547ag356D}
}

@ARTICLE{Roper2023,
       author = {{Roper}, Finn A. and {Oman}, Kyle A. and {Frenk}, Carlos S. and {Ben{\'\i}tez-Llambay}, Alejandro and {Navarro}, Julio F. and {Santos-Santos}, Isabel M.~E.},
        title = "{The diversity of rotation curves of simulated galaxies with cusps and cores}",
      journal = {\mnras},
         year = 2023,
        month = may,
       volume = {521},
       number = {1},
        pages = {1316-1336},
          doi = {10.1093/mnras/stad549},
archivePrefix = {arXiv},
       eprint = {2203.16652},
 primaryClass = {astro-ph.GA},
       adsurl = {https://ui.adsabs.harvard.edu/abs/2023MNRAS.521.1316R}
}

@ARTICLE{Pontzen2012,
       author = {{Pontzen}, Andrew and {Governato}, Fabio},
        title = "{How supernova feedback turns dark matter cusps into cores}",
      journal = {\mnras},
         year = 2012,
        month = apr,
       volume = {421},
       number = {4},
        pages = {3464-3471},
          doi = {10.1111/j.1365-2966.2012.20571.x},
archivePrefix = {arXiv},
       eprint = {1106.0499},
 primaryClass = {astro-ph.CO},
       adsurl = {https://ui.adsabs.harvard.edu/abs/2012MNRAS.421.3464P}
}

@ARTICLE{Read2005,
       author = {{Read}, J.~I. and {Gilmore}, G.},
        title = "{Mass loss from dwarf spheroidal galaxies: the origins of shallow dark matter cores and exponential surface brightness profiles}",
      journal = {\mnras},
         year = 2005,
        month = jan,
       volume = {356},
       number = {1},
        pages = {107-124},
          doi = {10.1111/j.1365-2966.2004.08424.x},
archivePrefix = {arXiv},
       eprint = {astro-ph/0409565},
 primaryClass = {astro-ph},
       adsurl = {https://ui.adsabs.harvard.edu/abs/2005MNRAS.356..107R}
}

@ARTICLE{Navarro1997,
       author = {{Navarro}, Julio F. and {Frenk}, Carlos S. and {White}, Simon D.~M.},
        title = "{A Universal Density Profile from Hierarchical Clustering}",
      journal = {\apj},
         year = 1997,
        month = dec,
       volume = {490},
       number = {2},
        pages = {493-508},
          doi = {10.1086/304888},
archivePrefix = {arXiv},
       eprint = {astro-ph/9611107},
 primaryClass = {astro-ph},
       adsurl = {https://ui.adsabs.harvard.edu/abs/1997ApJ...490..493N}
}

@ARTICLE{Navarro1996,
       author = {{Navarro}, Julio F. and {Frenk}, Carlos S. and {White}, Simon D.~M.},
        title = "{The Structure of Cold Dark Matter Halos}",
      journal = {\apj},
         year = 1996,
        month = may,
       volume = {462},
        pages = {563},
          doi = {10.1086/177173},
archivePrefix = {arXiv},
       eprint = {astro-ph/9508025},
 primaryClass = {astro-ph},
       adsurl = {https://ui.adsabs.harvard.edu/abs/1996ApJ...462..563N}
}

@ARTICLE{Garrison-Kimmel2014,
       author = {{Garrison-Kimmel}, Shea and {Boylan-Kolchin}, Michael and {Bullock}, James S. and {Kirby}, Evan N.},
        title = "{Too big to fail in the Local Group}",
      journal = {\mnras},
         year = 2014,
        month = oct,
       volume = {444},
       number = {1},
        pages = {222-236},
          doi = {10.1093/mnras/stu1477},
archivePrefix = {arXiv},
       eprint = {1404.5313},
 primaryClass = {astro-ph.GA},
       adsurl = {https://ui.adsabs.harvard.edu/abs/2014MNRAS.444..222G}
}

@ARTICLE{Boylan-Kolchin2012,
       author = {{Boylan-Kolchin}, Michael and {Bullock}, James S. and {Kaplinghat}, Manoj},
        title = "{The Milky Way's bright satellites as an apparent failure of {\ensuremath{\Lambda}}CDM}",
      journal = {\mnras},
         year = 2012,
        month = may,
       volume = {422},
       number = {2},
        pages = {1203-1218},
          doi = {10.1111/j.1365-2966.2012.20695.x},
archivePrefix = {arXiv},
       eprint = {1111.2048},
 primaryClass = {astro-ph.CO},
       adsurl = {https://ui.adsabs.harvard.edu/abs/2012MNRAS.422.1203B}
}

@ARTICLE{Boylan-Kolchin2011,
       author = {{Boylan-Kolchin}, Michael and {Bullock}, James S. and {Kaplinghat}, Manoj},
        title = "{Too big to fail? The puzzling darkness of massive Milky Way subhaloes}",
      journal = {\mnras},
         year = 2011,
        month = jul,
       volume = {415},
       number = {1},
        pages = {L40-L44},
          doi = {10.1111/j.1745-3933.2011.01074.x},
archivePrefix = {arXiv},
       eprint = {1103.0007},
 primaryClass = {astro-ph.CO},
       adsurl = {https://ui.adsabs.harvard.edu/abs/2011MNRAS.415L..40B}
}

@ARTICLE{Moore1999,
       author = {{Moore}, Ben and {Ghigna}, Sebastiano and {Governato}, Fabio and {Lake}, George and {Quinn}, Thomas and {Stadel}, Joachim and {Tozzi}, Paolo},
        title = "{Dark Matter Substructure within Galactic Halos}",
      journal = {\apjl},
         year = 1999,
        month = oct,
       volume = {524},
       number = {1},
        pages = {L19-L22},
          doi = {10.1086/312287},
archivePrefix = {arXiv},
       eprint = {astro-ph/9907411},
 primaryClass = {astro-ph},
       adsurl = {https://ui.adsabs.harvard.edu/abs/1999ApJ...524L..19M}
}

@ARTICLE{Klypin1999,
       author = {{Klypin}, Anatoly and {Kravtsov}, Andrey V. and {Valenzuela}, Octavio and {Prada}, Francisco},
        title = "{Where Are the Missing Galactic Satellites?}",
      journal = {\apj},
         year = 1999,
        month = sep,
       volume = {522},
       number = {1},
        pages = {82-92},
          doi = {10.1086/307643},
archivePrefix = {arXiv},
       eprint = {astro-ph/9901240},
 primaryClass = {astro-ph},
       adsurl = {https://ui.adsabs.harvard.edu/abs/1999ApJ...522...82K}
}

@ARTICLE{Burkert1995,
       author = {{Burkert}, A.},
        title = "{The Structure of Dark Matter Halos in Dwarf Galaxies}",
      journal = {\apjl},
         year = 1995,
        month = jul,
       volume = {447},
        pages = {L25-L28},
          doi = {10.1086/309560},
archivePrefix = {arXiv},
       eprint = {astro-ph/9504041},
 primaryClass = {astro-ph},
       adsurl = {https://ui.adsabs.harvard.edu/abs/1995ApJ...447L..25B}
}

@ARTICLE{Moore1994,
       author = {{Moore}, Ben},
        title = "{Evidence against dissipation-less dark matter from observations of galaxy haloes}",
      journal = {\nat},
         year = 1994,
        month = aug,
       volume = {370},
       number = {6491},
        pages = {629-631},
          doi = {10.1038/370629a0},
       adsurl = {https://ui.adsabs.harvard.edu/abs/1994Natur.370..629M}
}

@ARTICLE{Flores1994,
       author = {{Flores}, Ricardo A. and {Primack}, Joel R.},
        title = "{Observational and Theoretical Constraints on Singular Dark Matter Halos}",
      journal = {\apjl},
         year = 1994,
        month = may,
       volume = {427},
        pages = {L1},
          doi = {10.1086/187350},
archivePrefix = {arXiv},
       eprint = {astro-ph/9402004},
 primaryClass = {astro-ph},
       adsurl = {https://ui.adsabs.harvard.edu/abs/1994ApJ...427L...1F}
}

@ARTICLE{Bullock-BK2017,
       author = {{Bullock}, James S. and {Boylan-Kolchin}, Michael},
        title = "{Small-Scale Challenges to the {\ensuremath{\Lambda}}CDM Paradigm}",
      journal = {\araa},
         year = 2017,
        month = aug,
       volume = {55},
       number = {1},
        pages = {343-387},
          doi = {10.1146/annurev-astro-091916-055313},
archivePrefix = {arXiv},
       eprint = {1707.04256},
 primaryClass = {astro-ph.CO},
       adsurl = {https://ui.adsabs.harvard.edu/abs/2017ARA&A..55..343B}
}

@ARTICLE{Oman2015,
       author = {{Oman}, Kyle A. and {Navarro}, Julio F. and {Fattahi}, Azadeh and {Frenk}, Carlos S. and {Sawala}, Till and {White}, Simon D.~M. and {Bower}, Richard and {Crain}, Robert A. and {Furlong}, Michelle and {Schaller}, Matthieu and {Schaye}, Joop and {Theuns}, Tom},
        title = "{The unexpected diversity of dwarf galaxy rotation curves}",
      journal = {\mnras},
         year = 2015,
        month = oct,
       volume = {452},
       number = {4},
        pages = {3650-3665},
          doi = {10.1093/mnras/stv1504},
archivePrefix = {arXiv},
       eprint = {1504.01437},
 primaryClass = {astro-ph.GA},
       adsurl = {https://ui.adsabs.harvard.edu/abs/2015MNRAS.452.3650O}
}

@ARTICLE{DiCintio2014a,
       author = {{Di Cintio}, Arianna and {Brook}, Chris B. and {Macci{\`o}}, Andrea V. and {Stinson}, Greg S. and {Knebe}, Alexander and {Dutton}, Aaron A. and {Wadsley}, James},
        title = "{The dependence of dark matter profiles on the stellar-to-halo mass ratio: a prediction for cusps versus cores}",
      journal = {\mnras},
         year = 2014,
        month = jan,
       volume = {437},
       number = {1},
        pages = {415-423},
          doi = {10.1093/mnras/stt1891},
archivePrefix = {arXiv},
       eprint = {1306.0898},
 primaryClass = {astro-ph.CO},
       adsurl = {https://ui.adsabs.harvard.edu/abs/2014MNRAS.437..415D}
}

@ARTICLE{DiCintio2014b,
       author = {{Di Cintio}, Arianna and {Brook}, Chris B. and {Dutton}, Aaron A. and {Macci{\`o}}, Andrea V. and {Stinson}, Greg S. and {Knebe}, Alexander},
        title = "{A mass-dependent density profile for dark matter haloes including the influence of galaxy formation}",
      journal = {\mnras},
         year = 2014,
        month = jul,
       volume = {441},
       number = {4},
        pages = {2986-2995},
          doi = {10.1093/mnras/stu729},
archivePrefix = {arXiv},
       eprint = {1404.5959},
 primaryClass = {astro-ph.CO},
       adsurl = {https://ui.adsabs.harvard.edu/abs/2014MNRAS.441.2986D}
}

@ARTICLE{Tollet2016,
       author = {{Tollet}, Edouard and {Macci{\`o}}, Andrea V. and {Dutton}, Aaron A. and {Stinson}, Greg S. and {Wang}, Liang and {Penzo}, Camilla and {Gutcke}, Thales A. and {Buck}, Tobias and {Kang}, Xi and {Brook}, Chris and {Di Cintio}, Arianna and {Keller}, Ben W. and {Wadsley}, James},
        title = "{NIHAO - IV: core creation and destruction in dark matter density profiles across cosmic time}",
      journal = {\mnras},
         year = 2016,
        month = mar,
       volume = {456},
       number = {4},
        pages = {3542-3552},
          doi = {10.1093/mnras/stv2856},
archivePrefix = {arXiv},
       eprint = {1507.03590},
 primaryClass = {astro-ph.GA},
       adsurl = {https://ui.adsabs.harvard.edu/abs/2016MNRAS.456.3542T}
}

@ARTICLE{Garrison-Kimmel2019,
       author = {{Garrison-Kimmel}, Shea and {Hopkins}, Philip F. and {Wetzel}, Andrew and {Bullock}, James S. and {Boylan-Kolchin}, Michael and {Kere{\v{s}}}, Du{\v{s}}an and {Faucher-Gigu{\`e}re}, Claude-Andr{\'e} and {El-Badry}, Kareem and {Lamberts}, Astrid and {Quataert}, Eliot and {Sanderson}, Robyn},
        title = "{The Local Group on FIRE: dwarf galaxy populations across a suite of hydrodynamic simulations}",
      journal = {\mnras},
         year = 2019,
        month = jul,
       volume = {487},
       number = {1},
        pages = {1380-1399},
          doi = {10.1093/mnras/stz1317},
archivePrefix = {arXiv},
       eprint = {1806.04143},
 primaryClass = {astro-ph.GA},
       adsurl = {https://ui.adsabs.harvard.edu/abs/2019MNRAS.487.1380G}
}

@ARTICLE{Santos-Santos2020,
       author = {{Santos-Santos}, Isabel and {Navarro}, Julio F. and {Robertson}, Andrew and {Ben{\'\i}tez-Llambay}, Alejandro and {Oman}, Kyle A. and {Lovell}, Mark R. and {Frenk}, Carlos S. and {Ludlow}, Aaron D. and {Fattahi}, Azadeh and {Ritz}, Adam},
        title = "{Baryonic clues to the puzzling diversity of dwarf galaxy rotation curves}",
      journal = {\mnras},
         year = 2020,
        month = jun,
       volume = {495},
       number = {1},
        pages = {58-77},
          doi = {10.1093/mnras/staa1072},
archivePrefix = {arXiv},
       eprint = {1911.09116},
 primaryClass = {astro-ph.GA},
       adsurl = {https://ui.adsabs.harvard.edu/abs/2020MNRAS.495...58S}
}

@ARTICLE{Zentner2022,
       author = {{Zentner}, Aidan and {Dandavate}, Siddharth and {Slone}, Oren and {Lisanti}, Mariangela},
        title = "{A critical assessment of solutions to the galaxy diversity problem}",
      journal = {\jcap},
         year = 2022,
        month = jul,
       volume = {2022},
       number = {7},
          eid = {031},
        pages = {031},
          doi = {10.1088/1475-7516/2022/07/031},
archivePrefix = {arXiv},
       eprint = {2202.00012},
 primaryClass = {astro-ph.GA},
       adsurl = {https://ui.adsabs.harvard.edu/abs/2022JCAP...07..031Z}
}

@ARTICLE{Lelli2016,
       author = {{Lelli}, Federico and {McGaugh}, Stacy S. and {Schombert}, James M.},
        title = "{SPARC: Mass Models for 175 Disk Galaxies with Spitzer Photometry and Accurate Rotation Curves}",
      journal = {\aj},
         year = 2016,
        month = dec,
       volume = {152},
       number = {6},
          eid = {157},
        pages = {157},
          doi = {10.3847/0004-6256/152/6/157},
archivePrefix = {arXiv},
       eprint = {1606.09251},
 primaryClass = {astro-ph.GA},
       adsurl = {https://ui.adsabs.harvard.edu/abs/2016AJ....152..157L}
}

@ARTICLE{Cruz2025,
       author = {{Cruz}, Akaxia and {Brooks}, Alyson M. and {Lisanti}, Mariangela and {Peter}, Annika H.~G. and {Geda}, Robel and {Quinn}, Thomas and {Tremmel}, Michael and {Munshi}, Ferah and {Keller}, Ben and {Wadsley}, James},
        title = "{Galaxy Size and Rotation Curve Diversity in {\ensuremath{\Lambda}}CDM with Baryons}",
      journal = {\apj},
         year = 2026,
        month = jul,
       volume = {1005},
       number = {2},
          eid = {157},
        pages = {157},
          doi = {10.3847/1538-4357/ae69ce},
archivePrefix = {arXiv},
       eprint = {2510.11800},
 primaryClass = {astro-ph.GA},
       adsurl = {https://ui.adsabs.harvard.edu/abs/2026ApJ..1005..157C}
}

@ARTICLE{Brook2016b,
       author = {{Brook}, Chris B. and {Santos-Santos}, Isabel and {Stinson}, Greg},
        title = "{The different baryonic Tully-Fisher relations at low masses}",
      journal = {\mnras},
         year = 2016,
        month = jun,
       volume = {459},
       number = {1},
        pages = {638-645},
          doi = {10.1093/mnras/stw650},
archivePrefix = {arXiv},
       eprint = {1603.06595},
 primaryClass = {astro-ph.GA},
       adsurl = {https://ui.adsabs.harvard.edu/abs/2016MNRAS.459..638B}
}

@ARTICLE{Brook2016a,
       author = {{Brook}, Chris B. and {Shankar}, Francesco},
        title = "{A matter of measurement: rotation velocities and the velocity function of dwarf galaxies}",
      journal = {\mnras},
         year = 2016,
        month = feb,
       volume = {455},
       number = {4},
        pages = {3841-3847},
          doi = {10.1093/mnras/stv2550},
archivePrefix = {arXiv},
       eprint = {1506.00655},
 primaryClass = {astro-ph.CO},
       adsurl = {https://ui.adsabs.harvard.edu/abs/2016MNRAS.455.3841B}
}

@article{Vallat2018,
    title = {Pingouin: statistics in Python},
    volume = {3},
    DOI = {10.21105/joss.01026},
    number = {31},
    journal = {Journal of Open Source Software},
    publisher = {The Open Journal},
    author = {Vallat,  Raphael},
    year = {2018},
    month = nov,
    pages = {1026}
}

@ARTICLE{Santos-Santos2018,
       author = {{Santos-Santos}, Isabel and {Di Cintio}, Arianna and {Brook}, Chris B. and {Macci{\`o}}, Andrea and {Dutton}, Aaron and {Dom{\'\i}nguez-Tenreiro}, Rosa},
        title = "{NIHAO - XIV. Reproducing the observed diversity of dwarf galaxy rotation curve shapes in {\ensuremath{\Lambda}}CDM}",
      journal = {\mnras},
         year = 2018,
        month = feb,
       volume = {473},
       number = {4},
        pages = {4392-4403},
          doi = {10.1093/mnras/stx2660},
archivePrefix = {arXiv},
       eprint = {1706.04202},
 primaryClass = {astro-ph.GA},
       adsurl = {https://ui.adsabs.harvard.edu/abs/2018MNRAS.473.4392S}
}

@ARTICLE{Power2003,
       author = {{Power}, C. and {Navarro}, J.~F. and {Jenkins}, A. and {Frenk}, C.~S. and {White}, S.~D.~M. and {Springel}, V. and {Stadel}, J. and {Quinn}, T.},
        title = "{The inner structure of {\ensuremath{\Lambda}}CDM haloes - I. A numerical convergence study}",
      journal = {\mnras},
         year = 2003,
        month = jan,
       volume = {338},
       number = {1},
        pages = {14-34},
          doi = {10.1046/j.1365-8711.2003.05925.x},
archivePrefix = {arXiv},
       eprint = {astro-ph/0201544},
 primaryClass = {astro-ph},
       adsurl = {https://ui.adsabs.harvard.edu/abs/2003MNRAS.338...14P}
}

@ARTICLE{BrookDiCintio2015,
       author = {{Brook}, Chris B. and {Di Cintio}, Arianna},
        title = "{Signatures of dark matter halo expansion in galaxy populations}",
      journal = {\mnras},
         year = 2015,
        month = oct,
       volume = {453},
       number = {2},
        pages = {2133-2143},
          doi = {10.1093/mnras/stv1699},
archivePrefix = {arXiv},
       eprint = {1506.01019},
 primaryClass = {astro-ph.GA},
       adsurl = {https://ui.adsabs.harvard.edu/abs/2015MNRAS.453.2133B}
}

@ARTICLE{DiCintioLelli2016,
       author = {{Di Cintio}, Arianna and {Lelli}, Federico},
        title = "{The mass discrepancy acceleration relation in a {\ensuremath{\Lambda}}CDM context}",
      journal = {\mnras},
         year = 2016,
        month = feb,
       volume = {456},
       number = {1},
        pages = {L127-L131},
          doi = {10.1093/mnrasl/slv185},
archivePrefix = {arXiv},
       eprint = {1511.06616},
 primaryClass = {astro-ph.GA},
       adsurl = {https://ui.adsabs.harvard.edu/abs/2016MNRAS.456L.127D}
}

@ARTICLE{McGaugh2012,
       author = {{McGaugh}, Stacy S.},
        title = "{The Baryonic Tully-Fisher Relation of Gas-rich Galaxies as a Test of {\ensuremath{\Lambda}}CDM and MOND}",
      journal = {\aj},
         year = 2012,
        month = feb,
       volume = {143},
       number = {2},
          eid = {40},
        pages = {40},
          doi = {10.1088/0004-6256/143/2/40},
archivePrefix = {arXiv},
       eprint = {1107.2934},
 primaryClass = {astro-ph.CO},
       adsurl = {https://ui.adsabs.harvard.edu/abs/2012AJ....143...40M}
}

@ARTICLE{Girelli2020,
       author = {{Girelli}, G. and {Pozzetti}, L. and {Bolzonella}, M. and {Giocoli}, C. and {Marulli}, F. and {Baldi}, M.},
        title = "{The stellar-to-halo mass relation over the past 12 Gyr. I. Standard {\ensuremath{\Lambda}}CDM model}",
      journal = {\aap},
         year = 2020,
        month = feb,
       volume = {634},
          eid = {A135},
        pages = {A135},
          doi = {10.1051/0004-6361/201936329},
archivePrefix = {arXiv},
       eprint = {2001.02230},
 primaryClass = {astro-ph.CO},
       adsurl = {https://ui.adsabs.harvard.edu/abs/2020A&A...634A.135G}
}

@ARTICLE{Guo2010,
       author = {{Guo}, Qi and {White}, Simon and {Li}, Cheng and {Boylan-Kolchin}, Michael},
        title = "{How do galaxies populate dark matter haloes?}",
      journal = {\mnras},
         year = 2010,
        month = may,
       volume = {404},
       number = {3},
        pages = {1111-1120},
          doi = {10.1111/j.1365-2966.2010.16341.x},
archivePrefix = {arXiv},
       eprint = {0909.4305},
 primaryClass = {astro-ph.CO},
       adsurl = {https://ui.adsabs.harvard.edu/abs/2010MNRAS.404.1111G}
}

@ARTICLE{Moster2013,
       author = {{Moster}, Benjamin P. and {Naab}, Thorsten and {White}, Simon D.~M.},
        title = "{Galactic star formation and accretion histories from matching galaxies to dark matter haloes}",
      journal = {\mnras},
         year = 2013,
        month = feb,
       volume = {428},
       number = {4},
        pages = {3121-3138},
          doi = {10.1093/mnras/sts261},
archivePrefix = {arXiv},
       eprint = {1205.5807},
 primaryClass = {astro-ph.CO},
       adsurl = {https://ui.adsabs.harvard.edu/abs/2013MNRAS.428.3121M}
}

@ARTICLE{Stinson2006,
       author = {{Stinson}, Greg and {Seth}, Anil and {Katz}, Neal and {Wadsley}, James and {Governato}, Fabio and {Quinn}, Tom},
        title = "{Star formation and feedback in smoothed particle hydrodynamic simulations - I. Isolated galaxies}",
      journal = {\mnras},
         year = 2006,
        month = dec,
       volume = {373},
       number = {3},
        pages = {1074-1090},
          doi = {10.1111/j.1365-2966.2006.11097.x},
archivePrefix = {arXiv},
       eprint = {astro-ph/0602350},
 primaryClass = {astro-ph},
       adsurl = {https://ui.adsabs.harvard.edu/abs/2006MNRAS.373.1074S}
}

@ARTICLE{Chabrier2003,
       author = {{Chabrier}, Gilles},
        title = "{Galactic Stellar and Substellar Initial Mass Function}",
      journal = {\pasp},
         year = 2003,
        month = jul,
       volume = {115},
       number = {809},
        pages = {763-795},
          doi = {10.1086/376392},
archivePrefix = {arXiv},
       eprint = {astro-ph/0304382},
 primaryClass = {astro-ph},
       adsurl = {https://ui.adsabs.harvard.edu/abs/2003PASP..115..763C}
}

@ARTICLE{Dutton2017,
       author = {{Dutton}, Aaron A. and {Obreja}, Aura and {Wang}, Liang and {Gutcke}, Thales A. and {Buck}, Tobias and {Udrescu}, Silviu M. and {Frings}, Jonas and {Stinson}, Gregory S. and {Kang}, Xi and {Macci{\`o}}, Andrea V.},
        title = "{NIHAO XII: galactic uniformity in a {\ensuremath{\Lambda}}CDM universe}",
      journal = {\mnras},
         year = 2017,
        month = jun,
       volume = {467},
       number = {4},
        pages = {4937-4950},
          doi = {10.1093/mnras/stx458},
archivePrefix = {arXiv},
       eprint = {1610.06375},
 primaryClass = {astro-ph.GA},
       adsurl = {https://ui.adsabs.harvard.edu/abs/2017MNRAS.467.4937D}
}

@ARTICLE{Maccio2012,
       author = {{Macci{\`o}}, A.~V. and {Stinson}, G. and {Brook}, C.~B. and {Wadsley}, J. and {Couchman}, H.~M.~P. and {Shen}, S. and {Gibson}, B.~K. and {Quinn}, T.},
        title = "{Halo Expansion in Cosmological Hydro Simulations: Toward a Baryonic Solution of the Cusp/Core Problem in Massive Spirals}",
      journal = {\apjl},
         year = 2012,
        month = jan,
       volume = {744},
       number = {1},
          eid = {L9},
        pages = {L9},
          doi = {10.1088/2041-8205/744/1/L9},
archivePrefix = {arXiv},
       eprint = {1111.5620},
 primaryClass = {astro-ph.CO},
       adsurl = {https://ui.adsabs.harvard.edu/abs/2012ApJ...744L...9M}
}

@ARTICLE{Brook2012,
       author = {{Brook}, C.~B. and {Stinson}, G. and {Gibson}, B.~K. and {Wadsley}, J. and {Quinn}, T.},
        title = "{MaGICC discs: matching observed galaxy relationships over a wide stellar mass range}",
      journal = {\mnras},
         year = 2012,
        month = aug,
       volume = {424},
       number = {2},
        pages = {1275-1283},
          doi = {10.1111/j.1365-2966.2012.21306.x},
archivePrefix = {arXiv},
       eprint = {1201.3359},
 primaryClass = {astro-ph.CO},
       adsurl = {https://ui.adsabs.harvard.edu/abs/2012MNRAS.424.1275B}
}

@ARTICLE{Stinson2013,
       author = {{Stinson}, G.~S. and {Brook}, C. and {Macci{\`o}}, A.~V. and {Wadsley}, J. and {Quinn}, T.~R. and {Couchman}, H.~M.~P.},
        title = "{Making Galaxies In a Cosmological Context: the need for early stellar feedback}",
      journal = {\mnras},
         year = 2013,
        month = jan,
       volume = {428},
       number = {1},
        pages = {129-140},
          doi = {10.1093/mnras/sts028},
archivePrefix = {arXiv},
       eprint = {1208.0002},
 primaryClass = {astro-ph.CO},
       adsurl = {https://ui.adsabs.harvard.edu/abs/2013MNRAS.428..129S}
}

@ARTICLE{HaardtMadau2012,
       author = {{Haardt}, Francesco and {Madau}, Piero},
        title = "{Radiative Transfer in a Clumpy Universe. IV. New Synthesis Models of the Cosmic UV/X-Ray Background}",
      journal = {\apj},
         year = 2012,
        month = feb,
       volume = {746},
       number = {2},
          eid = {125},
        pages = {125},
          doi = {10.1088/0004-637X/746/2/125},
       adsurl = {https://ui.adsabs.harvard.edu/abs/2012ApJ...746..125H}
}

@ARTICLE{Ferland1998,
       author = {{Ferland}, G.~J. and {Korista}, K.~T. and {Verner}, D.~A. and {Ferguson}, J.~W. and {Kingdon}, J.~B. and {Verner}, E.~M.},
        title = "{CLOUDY 90: Numerical Simulation of Plasmas and Their Spectra}",
      journal = {\pasp},
         year = 1998,
        month = jul,
       volume = {110},
       number = {749},
        pages = {761-778},
          doi = {10.1086/316190},
       adsurl = {https://ui.adsabs.harvard.edu/abs/1998PASP..110..761F}
}

@ARTICLE{Shen2010,
       author = {{Shen}, S. and {Wadsley}, J. and {Stinson}, G.},
        title = "{The enrichment of the intergalactic medium with adiabatic feedback - I. Metal cooling and metal diffusion}",
      journal = {\mnras},
         year = 2010,
        month = sep,
       volume = {407},
       number = {3},
        pages = {1581-1596},
          doi = {10.1111/j.1365-2966.2010.17047.x},
archivePrefix = {arXiv},
       eprint = {0910.5956},
 primaryClass = {astro-ph.CO},
       adsurl = {https://ui.adsabs.harvard.edu/abs/2010MNRAS.407.1581S}
}

@ARTICLE{Wang2015,
       author = {{Wang}, Liang and {Dutton}, Aaron A. and {Stinson}, Gregory S. and {Macci{\`o}}, Andrea V. and {Penzo}, Camilla and {Kang}, Xi and {Keller}, Ben W. and {Wadsley}, James},
        title = "{NIHAO project - I. Reproducing the inefficiency of galaxy formation across cosmic time with a large sample of cosmological hydrodynamical simulations}",
      journal = {\mnras},
         year = 2015,
        month = nov,
       volume = {454},
       number = {1},
        pages = {83-94},
          doi = {10.1093/mnras/stv1937},
archivePrefix = {arXiv},
       eprint = {1503.04818},
 primaryClass = {astro-ph.GA},
       adsurl = {https://ui.adsabs.harvard.edu/abs/2015MNRAS.454...83W}
}

@PHDTHESIS{Stadel2001,
       author = {{Stadel}, Joachim Gerhard},
        title = "{Cosmological N-body simulations and their analysis}",
       school = {University of Washington, Seattle},
         year = 2001,
        month = jan,
       adsurl = {https://ui.adsabs.harvard.edu/abs/2001PhDT........21S}
}

@ARTICLE{Planck2014,
       author = {{Planck Collaboration} and {Ade}, P.~A.~R. and {Aghanim}, N. and {Armitage-Caplan}, C. and {Arnaud}, M. and {Ashdown}, M. and {Atrio-Barandela}, F. and {Aumont}, J. and {Baccigalupi}, C. and {Banday}, A.~J. and {Barreiro}, R.~B. and {Bartlett}, J.~G. and {Battaner}, E. and {Benabed}, K. and {Beno{\^\i}t}, A. and {Benoit-L{\'e}vy}, A. and {Bernard}, J.-P. and {Bersanelli}, M. and {Bielewicz}, P. and {Bobin}, J. and {Bock}, J.~J. and {Bonaldi}, A. and {Bond}, J.~R. and {Borrill}, J. and {Bouchet}, F.~R. and {Bridges}, M. and {Bucher}, M. and {Burigana}, C. and {Butler}, R.~C. and {Calabrese}, E. and {Cappellini}, B. and {Cardoso}, J.-F. and {Catalano}, A. and {Challinor}, A. and {Chamballu}, A. and {Chary}, R.-R. and {Chen}, X. and {Chiang}, H.~C. and {Chiang}, L.-Y. and {Christensen}, P.~R. and {Church}, S. and {Clements}, D.~L. and {Colombi}, S. and {Colombo}, L.~P.~L. and {Couchot}, F. and {Coulais}, A. and {Crill}, B.~P. and {Curto}, A. and {Cuttaia}, F. and {Danese}, L. and {Davies}, R.~D. and {Davis}, R.~J. and {de Bernardis}, P. and {de Rosa}, A. and {de Zotti}, G. and {Delabrouille}, J. and {Delouis}, J.-M. and {D{\'e}sert}, F.-X. and {Dickinson}, C. and {Diego}, J.~M. and {Dolag}, K. and {Dole}, H. and {Donzelli}, S. and {Dor{\'e}}, O. and {Douspis}, M. and {Dunkley}, J. and {Dupac}, X. and {Efstathiou}, G. and {Elsner}, F. and {En{\ss}lin}, T.~A. and {Eriksen}, H.~K. and {Finelli}, F. and {Forni}, O. and {Frailis}, M. and {Fraisse}, A.~A. and {Franceschi}, E. and {Gaier}, T.~C. and {Galeotta}, S. and {Galli}, S. and {Ganga}, K. and {Giard}, M. and {Giardino}, G. and {Giraud-H{\'e}raud}, Y. and {Gjerl{\o}w}, E. and {Gonz{\'a}lez-Nuevo}, J. and {G{\'o}rski}, K.~M. and {Gratton}, S. and {Gregorio}, A. and {Gruppuso}, A. and {Gudmundsson}, J.~E. and {Haissinski}, J. and {Hamann}, J. and {Hansen}, F.~K. and {Hanson}, D. and {Harrison}, D. and {Henrot-Versill{\'e}}, S. and {Hern{\'a}ndez-Monteagudo}, C. and {Herranz}, D. and {Hildebrandt}, S.~R. and {Hivon}, E. and {Hobson}, M. and {Holmes}, W.~A. and {Hornstrup}, A. and {Hou}, Z. and {Hovest}, W. and {Huffenberger}, K.~M. and {Jaffe}, A.~H. and {Jaffe}, T.~R. and {Jewell}, J. and {Jones}, W.~C. and {Juvela}, M. and {Keih{\"a}nen}, E. and {Keskitalo}, R. and {Kisner}, T.~S. and {Kneissl}, R. and {Knoche}, J. and {Knox}, L. and {Kunz}, M. and {Kurki-Suonio}, H. and {Lagache}, G. and {L{\"a}hteenm{\"a}ki}, A. and {Lamarre}, J.-M. and {Lasenby}, A. and {Lattanzi}, M. and {Laureijs}, R.~J. and {Lawrence}, C.~R. and {Leach}, S. and {Leahy}, J.~P. and {Leonardi}, R. and {Le{\'o}n-Tavares}, J. and {Lesgourgues}, J. and {Lewis}, A. and {Liguori}, M. and {Lilje}, P.~B. and {Linden-V{\o}rnle}, M. and {L{\'o}pez-Caniego}, M. and {Lubin}, P.~M. and {Mac{\'\i}as-P{\'e}rez}, J.~F. and {Maffei}, B. and {Maino}, D. and {Mandolesi}, N. and {Maris}, M. and {Marshall}, D.~J. and {Martin}, P.~G. and {Mart{\'\i}nez-Gonz{\'a}lez}, E. and {Masi}, S. and {Massardi}, M. and {Matarrese}, S. and {Matthai}, F. and {Mazzotta}, P. and {Meinhold}, P.~R. and {Melchiorri}, A. and {Melin}, J.-B. and {Mendes}, L. and {Menegoni}, E. and {Mennella}, A. and {Migliaccio}, M. and {Millea}, M. and {Mitra}, S. and {Miville-Desch{\^e}nes}, M.-A. and {Moneti}, A. and {Montier}, L. and {Morgante}, G. and {Mortlock}, D. and {Moss}, A. and {Munshi}, D. and {Murphy}, J.~A. and {Naselsky}, P. and {Nati}, F. and {Natoli}, P. and {Netterfield}, C.~B. and {N{\o}rgaard-Nielsen}, H.~U. and {Noviello}, F. and {Novikov}, D. and {Novikov}, I. and {O'Dwyer}, I.~J. and {Osborne}, S. and {Oxborrow}, C.~A. and {Paci}, F. and {Pagano}, L. and {Pajot}, F. and {Paladini}, R. and {Paoletti}, D. and {Partridge}, B. and {Pasian}, F. and {Patanchon}, G. and {Pearson}, D. and {Pearson}, T.~J. and {Peiris}, H.~V. and {Perdereau}, O. and {Perotto}, L. and {Perrotta}, F. and {Pettorino}, V. and {Piacentini}, F. and {Piat}, M. and {Pierpaoli}, E. and {Pietrobon}, D. and {Plaszczynski}, S. and {Platania}, P. and {Pointecouteau}, E.},
        title = "{Planck 2013 results. XVI. Cosmological parameters}",
      journal = {\aap},
         year = 2014,
        month = nov,
       volume = {571},
          eid = {A16},
        pages = {A16},
          doi = {10.1051/0004-6361/201321591},
archivePrefix = {arXiv},
       eprint = {1303.5076},
 primaryClass = {astro-ph.CO},
       adsurl = {https://ui.adsabs.harvard.edu/abs/2014A&A...571A..16P}
}

@ARTICLE{Wadsley2004,
       author = {{Wadsley}, J.~W. and {Stadel}, J. and {Quinn}, T.},
        title = "{Gasoline: a flexible, parallel implementation of TreeSPH}",
      journal = {\na},
         year = 2004,
        month = feb,
       volume = {9},
       number = {2},
        pages = {137-158},
          doi = {10.1016/j.newast.2003.08.004},
archivePrefix = {arXiv},
       eprint = {astro-ph/0303521},
 primaryClass = {astro-ph},
       adsurl = {https://ui.adsabs.harvard.edu/abs/2004NewA....9..137W}
}

@ARTICLE{Wadsley2017,
       author = {{Wadsley}, James W. and {Keller}, Benjamin W. and {Quinn}, Thomas R.},
        title = "{Gasoline2: a modern smoothed particle hydrodynamics code}",
      journal = {\mnras},
         year = 2017,
        month = oct,
       volume = {471},
       number = {2},
        pages = {2357-2369},
          doi = {10.1093/mnras/stx1643},
archivePrefix = {arXiv},
       eprint = {1707.03824},
 primaryClass = {astro-ph.IM},
       adsurl = {https://ui.adsabs.harvard.edu/abs/2017MNRAS.471.2357W}
}

@ARTICLE{Springel2010,
       author = {{Springel}, Volker},
        title = "{E pur si muove: Galilean-invariant cosmological hydrodynamical simulations on a moving mesh}",
      journal = {\mnras},
         year = 2010,
        month = jan,
       volume = {401},
       number = {2},
        pages = {791-851},
          doi = {10.1111/j.1365-2966.2009.15715.x},
archivePrefix = {arXiv},
       eprint = {0901.4107},
 primaryClass = {astro-ph.CO},
       adsurl = {https://ui.adsabs.harvard.edu/abs/2010MNRAS.401..791S}
}

@ARTICLE{Grand2017,
       author = {{Grand}, Robert J.~J. and {G{\'o}mez}, Facundo A. and {Marinacci}, Federico and {Pakmor}, R{\"u}diger and {Springel}, Volker and {Campbell}, David J.~R. and {Frenk}, Carlos S. and {Jenkins}, Adrian and {White}, Simon D.~M.},
        title = "{The Auriga Project: the properties and formation mechanisms of disc galaxies across cosmic time}",
      journal = {\mnras},
         year = 2017,
        month = may,
       volume = {467},
       number = {1},
        pages = {179-207},
          doi = {10.1093/mnras/stx071},
archivePrefix = {arXiv},
       eprint = {1610.01159},
 primaryClass = {astro-ph.GA},
       adsurl = {https://ui.adsabs.harvard.edu/abs/2017MNRAS.467..179G}
}

@ARTICLE{Sorce2015,
       author = {{Sorce}, Jenny G.},
        title = "{Minimization of biases in galaxy peculiar velocity catalogues}",
      journal = {\mnras},
         year = 2015,
        month = jul,
       volume = {450},
       number = {3},
        pages = {2644-2657},
          doi = {10.1093/mnras/stv760},
archivePrefix = {arXiv},
       eprint = {1504.06968},
 primaryClass = {astro-ph.CO},
       adsurl = {https://ui.adsabs.harvard.edu/abs/2015MNRAS.450.2644S}
}

@ARTICLE{Zaroubi1999,
       author = {{Zaroubi}, Saleem and {Hoffman}, Yehuda and {Dekel}, Avishai},
        title = "{Wiener Reconstruction of Large-Scale Structure from Peculiar Velocities}",
      journal = {\apj},
         year = 1999,
        month = aug,
       volume = {520},
       number = {2},
        pages = {413-425},
          doi = {10.1086/307473},
archivePrefix = {arXiv},
       eprint = {astro-ph/9810279},
 primaryClass = {astro-ph},
       adsurl = {https://ui.adsabs.harvard.edu/abs/1999ApJ...520..413Z}
}

@ARTICLE{Zaroubi1995,
       author = {{Zaroubi}, S. and {Hoffman}, Y. and {Fisher}, K.~B. and {Lahav}, O.},
        title = "{Wiener Reconstruction of the Large-Scale Structure}",
      journal = {\apj},
         year = 1995,
        month = aug,
       volume = {449},
        pages = {446},
          doi = {10.1086/176070},
archivePrefix = {arXiv},
       eprint = {astro-ph/9410080},
 primaryClass = {astro-ph},
       adsurl = {https://ui.adsabs.harvard.edu/abs/1995ApJ...449..446Z}
}

@ARTICLE{Hoffman1991,
       author = {{Hoffman}, Yehuda and {Ribak}, Erez},
        title = "{Constrained Realizations of Gaussian Fields: A Simple Algorithm}",
      journal = {\apjl},
         year = 1991,
        month = oct,
       volume = {380},
        pages = {L5},
          doi = {10.1086/186160},
       adsurl = {https://ui.adsabs.harvard.edu/abs/1991ApJ...380L...5H}
}

@ARTICLE{Bardeen1986,
       author = {{Bardeen}, J.~M. and {Bond}, J.~R. and {Kaiser}, N. and {Szalay}, A.~S.},
        title = "{The Statistics of Peaks of Gaussian Random Fields}",
      journal = {\apj},
         year = 1986,
        month = may,
       volume = {304},
        pages = {15},
          doi = {10.1086/164143},
       adsurl = {https://ui.adsabs.harvard.edu/abs/1986ApJ...304...15B}
}

@ARTICLE{Zeldovich1970,
       author = {{Zel'Dovich}, Ya. B.},
        title = "{Gravitational instability: An approximate theory for large density perturbations.}",
      journal = {\aap},
         year = 1970,
        month = mar,
       volume = {5},
        pages = {84-89},
       adsurl = {https://ui.adsabs.harvard.edu/abs/1970A&A.....5...84Z}
}

@ARTICLE{Tully2013,
       author = {{Tully}, R. Brent and {Courtois}, H{\'e}l{\`e}ne M. and {Dolphin}, Andrew E. and {Fisher}, J. Richard and {H{\'e}raudeau}, Philippe and {Jacobs}, Bradley A. and {Karachentsev}, Igor D. and {Makarov}, Dmitry and {Makarova}, Lidia and {Mitronova}, Sofia and {Rizzi}, Luca and {Shaya}, Edward J. and {Sorce}, Jenny G. and {Wu}, Po-Feng},
        title = "{Cosmicflows-2: The Data}",
      journal = {\aj},
         year = 2013,
        month = oct,
       volume = {146},
       number = {4},
          eid = {86},
        pages = {86},
          doi = {10.1088/0004-6256/146/4/86},
archivePrefix = {arXiv},
       eprint = {1307.7213},
 primaryClass = {astro-ph.CO},
       adsurl = {https://ui.adsabs.harvard.edu/abs/2013AJ....146...86T}
}

@ARTICLE{KnollmannKnebe2009,
       author = {{Knollmann}, Steffen R. and {Knebe}, Alexander},
        title = "{AHF: Amiga's Halo Finder}",
      journal = {\apjs},
         year = 2009,
        month = jun,
       volume = {182},
       number = {2},
        pages = {608-624},
          doi = {10.1088/0067-0049/182/2/608},
archivePrefix = {arXiv},
       eprint = {0904.3662},
 primaryClass = {astro-ph.CO},
       adsurl = {https://ui.adsabs.harvard.edu/abs/2009ApJS..182..608K}
}

@ARTICLE{Gill2014,
       author = {{Gill}, Stuart P.~D. and {Knebe}, Alexander and {Gibson}, Brad K.},
        title = "{The evolution of substructure - I. A new identification method}",
      journal = {\mnras},
         year = 2004,
        month = jun,
       volume = {351},
       number = {2},
        pages = {399-409},
          doi = {10.1111/j.1365-2966.2004.07786.x},
archivePrefix = {arXiv},
       eprint = {astro-ph/0404258},
 primaryClass = {astro-ph},
       adsurl = {https://ui.adsabs.harvard.edu/abs/2004MNRAS.351..399G}
}

@ARTICLE{Libeskind2020,
       author = {{Libeskind}, Noam I. and {Carlesi}, Edoardo and {Grand}, Robert J.~J. and {Khalatyan}, Arman and {Knebe}, Alexander and {Pakmor}, Ruediger and {Pilipenko}, Sergey and {Pawlowski}, Marcel S. and {Sparre}, Martin and {Tempel}, Elmo and {Wang}, Peng and {Courtois}, H{\'e}l{\`e}ne M. and {Gottl{\"o}ber}, Stefan and {Hoffman}, Yehuda and {Minchev}, Ivan and {Pfrommer}, Christoph and {Sorce}, Jenny G. and {Springel}, Volker and {Steinmetz}, Matthias and {Tully}, R. Brent and {Vogelsberger}, Mark and {Yepes}, Gustavo},
        title = "{The HESTIA project: simulations of the Local Group}",
      journal = {\mnras},
         year = 2020,
        month = oct,
       volume = {498},
       number = {2},
        pages = {2968-2983},
          doi = {10.1093/mnras/staa2541},
archivePrefix = {arXiv},
       eprint = {2008.04926},
 primaryClass = {astro-ph.GA},
       adsurl = {https://ui.adsabs.harvard.edu/abs/2020MNRAS.498.2968L}
}

@ARTICLE{Sorce2016,
       author = {{Sorce}, Jenny G. and {Gottl{\"o}ber}, Stefan and {Yepes}, Gustavo and {Hoffman}, Yehuda and {Courtois}, Helene M. and {Steinmetz}, Matthias and {Tully}, R. Brent and {Pomar{\`e}de}, Daniel and {Carlesi}, Edoardo},
        title = "{Cosmicflows Constrained Local UniversE Simulations}",
      journal = {\mnras},
         year = 2016,
        month = jan,
       volume = {455},
       number = {2},
        pages = {2078-2090},
          doi = {10.1093/mnras/stv2407},
archivePrefix = {arXiv},
       eprint = {1510.04900},
 primaryClass = {astro-ph.CO},
       adsurl = {https://ui.adsabs.harvard.edu/abs/2016MNRAS.455.2078S}
}

@ARTICLE{Carlesi2016,
       author = {{Carlesi}, Edoardo and {Sorce}, Jenny G. and {Hoffman}, Yehuda and {Gottl{\"o}ber}, Stefan and {Yepes}, Gustavo and {Libeskind}, Noam I. and {Pilipenko}, Sergey V. and {Knebe}, Alexander and {Courtois}, H{\'e}l{\`e}ne and {Tully}, R. Brent and {Steinmetz}, Matthias},
        title = "{Constrained Local UniversE Simulations: a Local Group factory}",
      journal = {\mnras},
         year = 2016,
        month = may,
       volume = {458},
       number = {1},
        pages = {900-911},
          doi = {10.1093/mnras/stw357},
archivePrefix = {arXiv},
       eprint = {1602.03919},
 primaryClass = {astro-ph.CO},
       adsurl = {https://ui.adsabs.harvard.edu/abs/2016MNRAS.458..900C}
}

@ARTICLE{Libeskind2010,
       author = {{Libeskind}, Noam I. and {Yepes}, Gustavo and {Knebe}, Alexander and {Gottl{\"o}ber}, Stefan and {Hoffman}, Yehuda and {Knollmann}, Steffen R.},
        title = "{Constrained simulations of the Local Group: on the radial distribution of substructures}",
      journal = {\mnras},
         year = 2010,
        month = jan,
       volume = {401},
       number = {3},
        pages = {1889-1897},
          doi = {10.1111/j.1365-2966.2009.15766.x},
archivePrefix = {arXiv},
       eprint = {0909.4423},
 primaryClass = {astro-ph.CO},
       adsurl = {https://ui.adsabs.harvard.edu/abs/2010MNRAS.401.1889L}
}

@ARTICLE{Gottloeber2010,
       author = {{Gottloeber}, Stefan and {Hoffman}, Yehuda and {Yepes}, Gustavo},
        title = "{Constrained Local UniversE Simulations (CLUES)}",
      journal = {arXiv e-prints},
         year = 2010,
        month = may,
          eid = {arXiv:1005.2687},
        pages = {arXiv:1005.2687},
          doi = {10.48550/arXiv.1005.2687},
archivePrefix = {arXiv},
       eprint = {1005.2687},
 primaryClass = {astro-ph.CO},
       adsurl = {https://ui.adsabs.harvard.edu/abs/2010arXiv1005.2687G}
}

@ARTICLE{Waterval2024,
       author = {{Waterval}, Stefan and {Macci{\`o}}, Andrea V. and {Buck}, Tobias and {Obreja}, Aura and {Cho}, Changhyun and {Jin}, Zehao and {Davis}, Benjamin L. and {Dixon}, Keri L. and {Kang}, Xi},
        title = "{HELLO project: high-z evolution of large and luminous objects}",
      journal = {\mnras},
         year = 2024,
        month = sep,
       volume = {533},
       number = {2},
        pages = {1463-1484},
          doi = {10.1093/mnras/stae1913},
archivePrefix = {arXiv},
       eprint = {2403.03985},
 primaryClass = {astro-ph.GA},
       adsurl = {https://ui.adsabs.harvard.edu/abs/2024MNRAS.533.1463W}
}

@ARTICLE{Blank2019,
       author = {{Blank}, Marvin and {Macci{\`o}}, Andrea V. and {Dutton}, Aaron A. and {Obreja}, Aura},
        title = "{NIHAO - XXII. Introducing black hole formation, accretion, and feedback into the NIHAO simulation suite}",
      journal = {\mnras},
         year = 2019,
        month = aug,
       volume = {487},
       number = {4},
        pages = {5476-5489},
          doi = {10.1093/mnras/stz1688},
archivePrefix = {arXiv},
       eprint = {1906.06955},
 primaryClass = {astro-ph.GA},
       adsurl = {https://ui.adsabs.harvard.edu/abs/2019MNRAS.487.5476B}
}

\appendix

\section{Alternative rotation curve measurements}  \label{appendix:RCs}

\begin{figure*}[!t]
\centering
  \includegraphics[width=13cm]{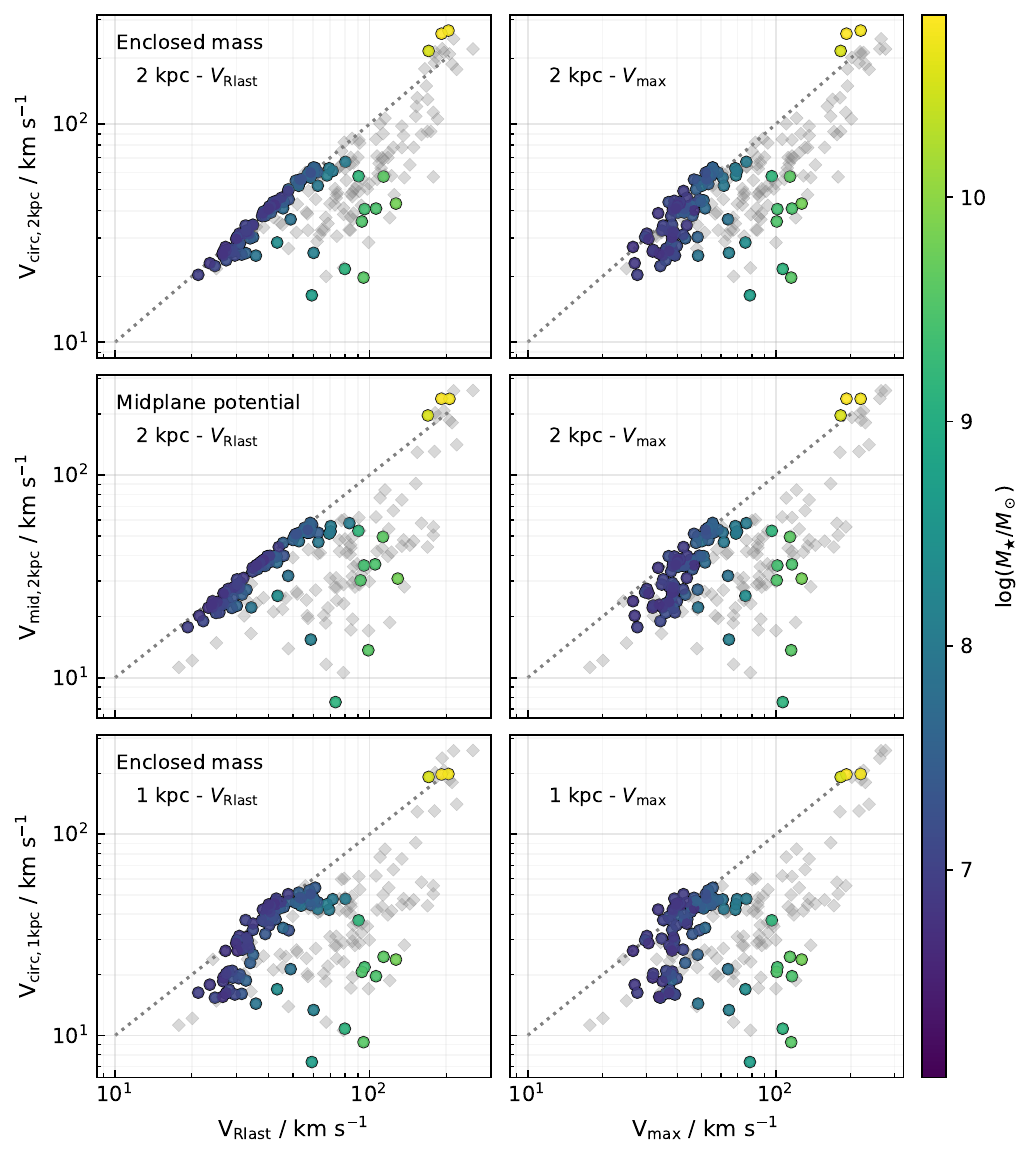}
  \caption{Relation between inner and outer rotation-curve velocities for the \texttt{NIVARIA-LG} simulated galaxy sample (circles) and the SPARC observations (grey diamonds), analogous to \Fig{fig:RCs-comparison}, but using different methods for measuring the rotation curves and different characteristic radii. In the first row, the enclosed-mass method is used, with the inner velocity measured at 2~kpc and the outer velocity measured at $R_\mathrm{last}$ (left) or $V_\mathrm{max}$ (right). The second row shows the same measurements using the midplane potential method. The third row uses the enclosed-mass method with the inner velocity measured at 1~kpc.}
  \label{fig:RC-appendix}
\end{figure*}

We test the robustness of our rotation curve measurements to the choice of characteristic radii and to the method used to compute the circular velocity. \Fig{fig:RC-appendix} compares the inner and outer circular velocities for the \texttt{NIVARIA-LG} galaxies and the SPARC observations, analogous to \Fig{fig:RCs-comparison}, using three different measurement choices. The first row uses our fiducial enclosed-mass method, with the inner velocity measured at 2~kpc; the second row uses the midplane-potential method, in which the circular velocity is computed directly from the radial gravitational acceleration in the disc midplane, with the same 2~kpc inner radius; and the third row uses the enclosed-mass method with the inner velocity measured at 1~kpc. In the left column, the outer velocity is measured at $R_\mathrm{last}$, while in the right column it is taken as the maximum circular velocity, $V_\mathrm{max}$.

The different choices produce qualitatively similar distributions for both the simulated and observed galaxies, indicating that the broad diversity of rotation curve shapes is not strongly dependent on the adopted measurement. In particular, the midplane-potential method gives a similar overall relation to the enclosed-mass calculation, while changing the inner radius from 2 to 1~kpc or replacing $R_\mathrm{last}$ with $V_\mathrm{max}$ does not qualitatively alter the distribution. We therefore retain $V_\mathrm{2kpc}/V_\mathrm{Rlast}$ as our fiducial definition throughout the paper.

\section{HESTIA intermediate resolution}

To assess the impact of numerical resolution on our comparison with \texttt{HESTIA}, we repeat the analysis shown in \Fig{fig:RCs-comparison} using the intermediate-resolution \texttt{HESTIA} simulations, which match the resolution of \texttt{NIVARIA-LG}. While the main text uses the high-resolution \texttt{HESTIA} simulations, this comparison allows us to test whether the differences between the two simulation suites arise from their different resolutions. As shown in \Fig{fig:RCs-comparison}, the overall distribution of rotation curve shapes remains similar to that obtained with the high-resolution \texttt{HESTIA} simulations. In particular, the differences between \texttt{NIVARIA-LG} and \texttt{HESTIA} persist at matched resolution, indicating that they are not driven  by different  numerical resolution.

\begin{figure}
\centering
  \includegraphics[width=8cm]{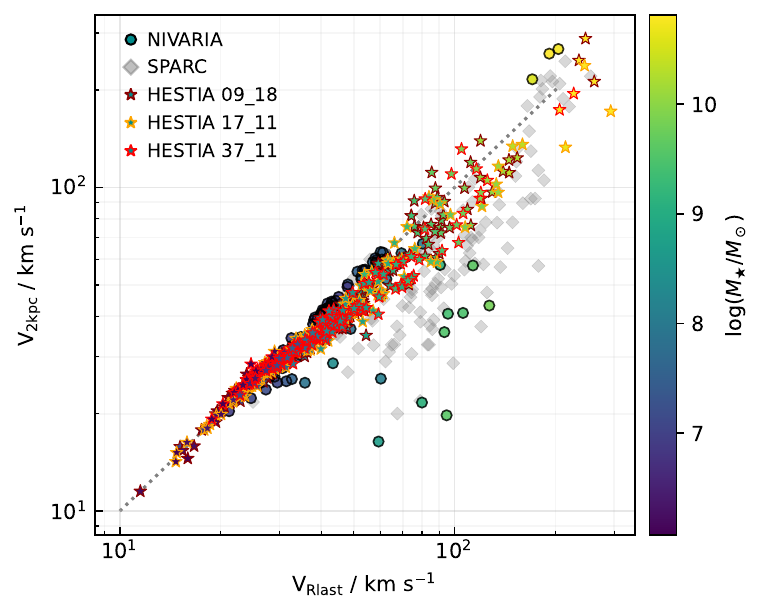}
  \caption{$V_\mathrm{2kpc}-V_\mathrm{Rlast}$ relation for the \texttt{NIVARIA-LG} simulated galaxy sample (circles), SPARC observations (grey diamonds), and the three intermediate-resolution \texttt{HESTIA} simulations, following the same format as the bottom-left panel of \Fig{fig:RCs-comparison}. All simulated galaxies are coloured by their stellar mass.}
\label{fig:RCs-comparison}
\end{figure}


\label{lastpage}
\end{document}